# Polynomial Time Algorithm for Boolean Satisfiability Problem

Stepan G. Margaryan

*e-mail: stmargaryan@gmail.com*
*Yerevan, Armenia*

## Abstract:

This is the latest in a series of articles aimed at exploring the relationship between the complexity classes of $P$ and $NP$.

In the previous papers, we have proved that the *sat CNF* problem is polynomially reduced to the problem of finding a special covering for a set under the special decomposition of this set and vice versa. That is, these problems are polinomially equivalent. This means that the problem of finding a special covering for a set under a special decomposition of this set, is an $NP$-complete problem.

We also described algorithmic procedures that determine whether there is a special covering for a set under a special decomposition of this set.

In this article we prove that all these algorithmic procedures have polynomial time complexity with respect to the length of input data.

In addition, we will describe an algorithm that, given any Boolean function in conjunctive normal form ($CNF$), determines in polynomial time whether this function is satisfiable. We will prove that the time complexity of this algorithm is bounded by the cube of the length of the input data. Also, if the function is not satisfiable, the algorithm deduces this result noting the reason for this result.

We have implemented an algorithm in Python, and successfully tested it on Boolean functions represented in $CNF$ with tens of thousands of variables and tens of thousands of clauses.

## 1. Introduction

The search for algorithms with polynomial running time complexity for determining the satisfiability of propositional formulas began long before the notion of $NP$-complete problems appeared. This is so as algorithms also play significant role in the procedures of many proofs of propositional formulas.

Various types of algorithms have been created. Each type gave good results in terms of time complexity when applied to formulas with certain restrictions.

In [9], [11], [12], [13] widely known probabilistic algorithms are described and a comprehensive analysis of estimates of their complexity is carried out.

However, it turns out that basically, the ''behavior'' of the algorithm strictly depends on the probability of the distribution of literals in clauses.

Many authors use also a procedure which eliminates specific clauses such that the satisfiability of the formula under consideration is preserved. In [14], the clause elimination procedure and its possible effectiveness on time complexity are widely studied and analyzed.

The algorithms based on any known concept, for finding the satisfiability of Boolean functions, are presented in [15]. Any type of these algorithms based on some notion is well-studied and also a comprehensive analysis is carried out. In addition, specific practical algorithms are also studied.

Although the emergence of the theory of $NP$-complete problems ([1], [4], [6]) created the reliable basis for a huge theoretical breakthrough in the theory of algorithms, the situation in the field of practical algorithms for solving $NP$-complete problems, is almost the same. In many cases, the same algorithm that gives a good result when applying it to the formula under some restrictions, may behave uncertainty in applying to other formulas.

In this article, we describe a new algorithm for finding the satisfiability of a Boolean function based on new concepts called ''special decomposition of a set'' and ''special covering for a set under the special decomposition of this set'' introduced in [7], [8].

Based on the results of [8], we will show that this new algorithm is applicable to any Boolean function represented in conjunctive normal form. And also, the running time estimate for this algorithm is a polynomial, which does not depend on the number of literals in individual clauses, but depends only on the total number of literals in all clauses.

In addition, if the given function is satisfiable, then the algorithm outputs the corresponding values of the variables in polynomial time. If the function is not satisfiable, the algorithm deduces this result noting the reason for that result, also in polynomial time.

The idea of studying the notions of a special decomposition of a set and a special covering for a set under the special decomposition of this set arose as a result of the following observations:

The set of clauses of a Boolean function represented in $CNF$ can be decomposed into subsets, in such a manner, that each subset contains only the clauses that include the same specific literal. With such a decomposition of the set of clauses, for any variable of the function, two non-intersecting subsets will be formed. At the same time, since any variable of the function is included in some clause, either as a positive literal or a negative literal, at least one of these subsets is necessarily non-empty.

Obviously, all clauses included in the same subset takes the value 1 if the literal representing this subset takes the value 1. Therefore, if it is possible to find a set of subsets such that each of them corresponds to different variables of the function (either a positive literal or a negative literal), and all together cover the entire set of clauses, then it is easy to find an assignment of variables that ensures the satisfiability of this function.

Thus, the problem of finding out whether a function represented in conjunctive normal form is satisfiable reduces to the problem of finding corresponding subsets of the set of clauses such that they will cover the set of all clauses.

In [7], [8], we introduced the concepts of a special decomposition of a set and a special covering for a set under a special decomposition of this set.

The purpose of these concepts was, in fact, to search for conditions under which the following is possible:

- given a non-empty set and ordered pairs of non-intersecting subsets of this set, select one component from each of the ordered pairs in a way, that the selected subsets cover the set. We called it a special covering.

In [8] we have proved that there are necessary and sufficient conditions for the existence of a special covering for a set under the given special decomposition.

However, verification of any of these conditions requires some algorithmic procedure.

Note that, all of these procedures are described in [8] in the course of their respective proofs. This implies that there exists a deterministic procedure which, for any given special decomposition of a set, can either find the special covering for that set (if it exists), or deduce that there is no special covering under the given special decomposition.

In [7], [8] important results have been obtained concerning the decidability of the Boolean satisfiability problem and the decidability of the problem of existence of a special covering for a set. We have proved that these problems are polinomially equivalent.

In particularly, in [8] it was proven that the Boolean function represented in conjunctive normal form is satisfiable if and only if there is a special covering for the set of clauses under the special decomposition of this set.

Necessary and sufficient conditions for Boolean satisfiability were obtained in [8], when applying such a procedure on the special decomposition generated by a Boolean function represented in $CNF$.

We will describe partial algorithms that perform all of these procedures. We also estimate the number of elementary operations required by these algorithms.

As a result, the algorithm for searching a special covering for the given set is actually the union of these partial algorithms.

We also describe a transition algorithm which, based on the input data for Boolean function in $CNF$, outputs the input data for the algorithm of searching a special covering for the set of clauses of this function. Thus, we get an algorithm for determining the satisfiability of the given Boolean function.

The significant difference between this algorithm and existing ones is as follows:

- the algorithm gives a result for any Boolean function in $CNF$,

- the algorithm only works as a deterministic procedure without any random steps.

- the time complexity of this algorithm is bounded by the number $(N(f))^3$, where $N(f)$ is the length of the input data of the given Boolean function represented in $CNF$.

## 2. Basic descriptions

Let's recall some basic definitions and notation from [7] and [8].

$S = \{e_1, e_2, \ldots, e_m\}$ is a non-empty set of $m$ elements for some natural number $m$.

We assume that, $n$ arbitrary ordered pairs of arbitrarily subsets of the set $S$ are given, for some natural number $n$. In [8] we used the notation $(M_i^\alpha, M_i^{1-\alpha})$ for these ordered pairs, where $\alpha \in \{0,1\}$ and $i \in \{1, \ldots, n\}$.

The subsets $M_i^\alpha$ and $M_i^{1-\alpha}$ are called the $\alpha$-component and $(1-\alpha)$-component of the ordered pair $(M_i^\alpha, M_i^{1-\alpha})$, respectively.

Recall that $d_nS$ is an arbitrarily ordered set of arbitrarily ordered pairs of those subsets:

$$d_nS = \{(M_1^\alpha, \ M_1^{1-\alpha}), (M_2^\alpha, \ M_2^{1-\alpha}), \ . \ . \ . \ , (M_n^\alpha, \ M_n^{1-\alpha})\}.$$

<u>*Definition*</u> 2.1 The set $d_nS$ will be called a special decomposition of the set $S$, if

        2.1.1. $\forall \ i \in \{1, \ldots, n\}$ $(M_i^\alpha \cap M_i^{1-\alpha}) = \emptyset$,

        2.1.2. $\forall \ i \in \{1, \ldots, n\}$ $(M_i^\alpha \neq \emptyset)$ or $(M_i^{1-\alpha} \neq \emptyset)$,

        2.1.3. $\bigcup_{i=1}^n (M_i^\alpha \cup M_i^{1-\alpha}) = S$.

<u>*Definition*</u> 2.2. Let the set $d_nS$ be a special decomposition of the set $S$.

For some $\alpha_1, \alpha_2, \ldots, \alpha_n$, where $\alpha_i \in \{0,1\}$, the ordered set

$$c_nS = \{M_1^{\alpha_1}, M_2^{\alpha_2}, \ldots, M_n^{\alpha_n}\},$$

will be called a special covering for the set $S$ under the special decomposition $d_nS$, if

$$\bigcup_{i=1}^n M_i^{\alpha_i} = S.$$

Let us recall some notations from [7] and [8], which we will also use in this article.

For a special decomposition $d_nS$,

$$d_nS = \{(M_1^\alpha, M_1^\alpha), \ldots, (M_n^\alpha, M_n^{1-\alpha})\}$$

and for an $\alpha \in \{0, 1\}$, we use the following notation:

$sM^\alpha = \{M_1^\alpha, M_2^\alpha, \ldots, M_n^\alpha\}$. It is called the set of subsets of the $\alpha$-domain of the decomposition $d_nS$ or, briefly, the $\alpha$-domain of the decomposition $d_nS$.

$M^\alpha = \bigcup_{i=1}^n M_i^\alpha$. The set $M^\alpha$ is called a set of elements of the $\alpha$-domain.

$(i_1, \ldots, i_k)I(d_nS)$ is the set of ordered pairs obtained by permuting the components of the ordered pairs

$$\{(M_{i_1}^\alpha, \ M_{i_1}^{1-\alpha}), \ldots, (M_{i_l}^\alpha, \ M_{i_l}^{1-\alpha})\} \subseteq d_nS.$$

Sometimes, if no ambiguity arises, we will skip the index numbering and use the short notation $I(d_nS)$. It is easy to see that $(i_1, \ldots, i_k)I(d_nS)$ is a special decomposition.

$(i_1, \ldots, i_k)sM^\alpha$ is the set obtained by placing the subsets $M_{i_1}^{1-\alpha}, \ldots, M_{i_k}^{1-\alpha}$ in the set $sM^\alpha$ instead of subsets $M_{i_1}^\alpha, \ldots, M_{i_k}^\alpha$, respectively.

$(i_1, \ldots, i_k)sM^\alpha$ is called the $\alpha$-domain of the decomposition $(i_1, \ldots, i_k)I(d_nS)$.

*If under the special decomposition $d_nS$, the set $sM^\alpha$ is a special covering for the set $S$ for some $\alpha \in \{0, 1\}$, then such a covering will be called a special $M^\alpha$-covering for the set $S$.*

If the Boolean function $f(x_1, \ldots, x_n)$ is represented in $CNF$ with $m$ clauses, then denote by $S(f)$ the set of clauses of this function:

$$S(f) = \{c_1, c_2, \ldots, c_m\}.$$

For each $i \in \{1, \ldots, n\}$ and $\alpha \in \{0,1\}$ the sets $fM_i^{\alpha}$ and $fM_i^{1-\alpha}$ are composed as follows:

$$fM_i^{\alpha} = \{c_j \, / \, c_j \in S(f) \text{ and } c_j \text{ contains the literal } x_i^{\alpha}, \ (j \in \{1, \ldots, m\})\}.$$
$$fM_i^{1-\alpha} = \{c_j \, / \, c_j \in S(f) \text{ and } c_j \text{ contains the literal } x_i^{1-\alpha}, \ (j \in \{1, \ldots, m\})\}.$$

The set

$$d_n S(f) = \{(fM_1^{\alpha}, fM_1^{1-\alpha}), (fM_2^{\alpha}, fM_2^{1-\alpha}), \ldots, (fM_n^{\alpha}, fM_n^{1-\alpha})\}$$

is an ordered set of ordered pairs of these subsets.

Recall some important results from [8].

According to Lemma 2.5 from [8], there exists a special covering for the set $S$ under the decomposition $d_n S$ if and only if for some $\alpha \in \{0,1\}$ there exists an $M^{\alpha}$-covering for $S$ under some decomposition $I(d_n S)$.

According to Theorem 6.13 in [8], there are necessary and sufficient conditions for the existence of a special covering for a set, that is:

There is a special $M^{\alpha}$-covering for the set $S$ under the special decomposition $d_n S$ if and only if the set of elements not included in the $\alpha$-domain of the decomposition $d_n S$ and the graph extension elements are stable with respect to both the cleaning procedure and the compatibility procedure in the extended graph.

In addition, according to Theorems 8.1 and 8.4 in [8], similar conditions lead to necessary and sufficient conditions for establishing the satisfiability of a Boolean function in conjunctive normal form.

The proofs of these theorems are based on the concepts of a replaceable subset and a replaceability procedure, and the properties of pointing graphs [8] are also used.

At the same time, as a result of the procedures used to proof these theorems, we obtain a set of subsets, which is a special covering for the set under consideration, if such subsets exist. Also, if such subsets do not exist, as a result of the mentioned procedures we will get this conclusion.

Also, considering the special decomposition corresponding to a function $f(x_1, \ldots, x_n)$ represented in conjunctive normal form, as a result of the same procedure, we obtain the Boolean values of the variables of this function for which the function takes the value 1, if such an assignment exists.

Using the results of these theorems, a general procedure for determining the satisfiability of a Boolean function in $CNF$ is described in [8], Section 9. We will investigate and estimate the number of elementary operations, required to complete the procedure.

According to Lemma 7.1 in [8],
$$d_n S(f) = \{(fM_1^\alpha, fM_1^{1-\alpha}), (fM_2^\alpha, fM_2^{1-\alpha}), \ldots, (fM_n^\alpha, fM_n^{1-\alpha})\}$$
is a special decomposition of the set $S(f)$.

According to Theorem 7.2 in [8], the function $f$ is satisfiable if and only if there is a special covering for the set $S(f)$ under the special decomposition $d_n S(f)$.

Therefore,

*we first consider the description of the general procedure for finding a special covering of a set under the special decomposition of this set.*

Assume we are given a non-empty set $S = \{e_1, \ldots, e_m\}$, for some natural number $m$. Also, for some natural number $n$, the set
$$d_n S = \{(M_1^\alpha, M_1^{1-\alpha}), (M_2^\alpha, M_2^{1-\alpha}), \ldots, (M_n^\alpha, M_n^{1-\alpha})\}$$
is a special decomposition of the set $S$.

Based on Section 9 in [8], we describe an algorithmic procedure to search for a special covering for the set $S$ under the special decomposition $d_n S$, the algorithm is called $SSC$.

### $SSC$. Search for Special Covering

1.  Find the set of elements not included in the $\alpha$-domain;   $// X = \{e_{q_1}, \ldots, e_{q_l}\} = S \setminus M^\alpha //$
2.  **If** $X = \emptyset$:
3.      **Output** ("$\alpha$-domain of the given decomposition is a special covering for the set $S$
4.      **Stop**;
5.  **Else**:
6.      Construct the pointing graph $G$ associated with elements $\{e_{q_1}, \ldots, e_{q_l}\}$
7.      Search for useless vertices in the graph

8.      **While** a useless vertex $v$ is found:
9.         Apply the *Removal Procedure* to $v$
10.        **If** $v$ is not a removable vertex:   //removable vertex is defined in p. 5.1.1 in [8] //
11.           **Output** "There is no special covering"
12.           **Stop**;
13.        **Else**:
14.           Repeat searching for another useless vertex
15.        **EndIf**;
16.     **EndWhile**;

17.     IncompSetIndicator := 0   // indicator showing the presence of incompatible sets //
18.     Search for incompatible set in the pointing graph

19.     **While** there exists an incompatible set with respect to some element $e \in S \setminus M^\alpha$:
20.        IncompSetIndicator := 1
21.        Apply the *Compatibility Procedure* to the found set

22.        **If** incompatibility of the set under consideration is eliminated:

23.         Repeat searching for incompatible set with respect to another element $e$

24.      **Else:**

25.        **If** none of the $(1-\alpha)$-components of ordered pairs not included in the replaceability procedure contains the element $e$:

26.          **Output** ("There is no special covering")    // $e$ is an unreachable element //

27.          **Stop;**

28.        **Else:**

29.          Add the element $e$ to the set $Y$    // $Y$ is the set of elements by which the considered graph will be extended //

30.          Repeat searching for incompatible set with respect to another element $e$

31.        **EndIf;**

32.      **EndIf;**

33.    **EndWhile;**

34.    **If** IncompSetIndicator = 0:

35.      **Output** "$\alpha$-domain is a special covering"

36.      **Stop;**

37.    **Else:**

38.      **If** $Y = \emptyset$:

39.        **Output** ("$\alpha$-domain is a special covering")

40.        **Stop;**

41.      **Else:**

42.        Add to the graph in question new main vertices associated with elements included in $Y$

43.        Go to *Constructing Procedure* to extended the graph

44.      **EndIf;**

45.    **EndIf;**

46.  **EndIf;**

As can be seen, the general procedure SSC consists of various distinct algorithmic components:

P1) The procedure for finding the elements that are not included in the $\alpha$-domain and constructing the pointing graph,

P2) The cleaning procedure of the pointing graph, if there are useless vertices in it.

P3) The incompatibility elimination procedure in the pointing graph, if there are incompatible sets of vertices in it.

P4) The extending procedure.

Our goal is to describe the algorithms for each of these components and estimate the number of operations required to complete the procedure, relative to the length of input data.

### 3. Data Representation

We will be dealing with a non-empty set of $m$ elements, denoted as

$$S = \{e_1, \ldots, e_m\}.$$

For technical convenience we consider $S$ as an ordered set with some natural numbering of its elements. It will not lead to any ambiguity.

Each time we use them, we will assume that they are numbered in the order in which they occur in the designation of a set. We do this in order to represent a special decomposition of the set $S$ by means of ordered pair of $(0,1)$ matrices.

Based on a special decomposition

$$d_n S = \{(M_1^\alpha, M_1^\alpha), \ldots, (M_n^\alpha, M_n^{1-\alpha})\},$$

we form $(n \times m)$ matrices denoted by $sM\alpha$ and $sM\bar{\alpha}$, as well as one-dimensional arrays $M\alpha$ and $M\bar{\alpha}$ of the size $m$, where $\bar{\alpha}=1$, if $\alpha=0$ and $\bar{\alpha}=0$, if $\alpha=1$.

The elements of these matrices and arrays are determined as follows:
For $i \in \{1, \ldots, n\}$ and $j \in \{1, \ldots, m\}$,

$$sM\alpha(i, j) = \begin{cases} 0, & if\ e_j \notin M_i^\alpha \\ 1, & if\ e_j \in M_i^\alpha \end{cases} \qquad sM\bar{\alpha}(i, j) = \begin{cases} 0, & if\ e_j \notin M_i^{1-\alpha} \\ 1, & if\ e_j \in M_i^{1-\alpha} \end{cases}.$$

For $j \in \{1, \ldots, m\}$,
- $M\alpha(j)$ is the number of 1s in the $j$-th column of the matrix $sM\alpha$.
- $M\bar{\alpha}(j)$ is the number of 1s in the $j$-th column of the matrix $sM\bar{\alpha}$.

It is easy to see that with such notation, for $j \in \{1, \ldots, m\}$, we will have the following:

$M\alpha(j)$ is the number of occurrences of the element $e_j$ in the subsets of the $\alpha$-domain of the given decomposition.

$M\bar{\alpha}(j)$ is the number of occurrences of the element $e_j$ in the subsets of the $(1-\alpha)$-domain of the given decomposition.

*We will say that the ordered pair of $(0,1)$ matrices $(sM\alpha, sM\bar{\alpha})$ corresponds to the special decomposition $d_n S$, if this pair is formed in described manner on the basis of the decomposition $d_n S$.*

On the other hand, the special decomposition $d_n S$ is determined by the corresponding ordered pair of matrices $(sM\alpha, sM\bar{\alpha})$, that is:

For any $i \in \{1, \ldots, n\}$ and $j \in \{1, \ldots, m\}$, the elements of the subsets $M_i^\alpha$ and $M_i^{1-\alpha}$ are uniquely determined by the non-zero elements of the $i$-th rows of the matrices $sM\alpha$ and $sM\bar{\alpha}$, respectively:

$$M_i^\alpha = \{e_j \in S\ /\ sM\alpha(i, j) = 1\},$$
$$M_i^{1-\alpha} = \{e_j \in S\ /\ sM\bar{\alpha}(i, j) = 1\},$$

The following statements are obvious:

- if $M^\alpha$ is the $\alpha$-domain of the given decomposition and $e_j \notin M^\alpha$, for some element $e_j \in S$, then the $j$-th column of the matrix $sM\alpha$ consists only of zero elements.

- for any $\alpha \in \{0, 1\}$, the non-zero elements of the $j$-th column of the matrix $sM\alpha$ determine the subsets of $\alpha$-domain containing the element $e_j$,

- the elements of the $\alpha$-domain, as well as the $(1-\alpha)$-domain are uniquely determined by the non-zero elements of the arrays $M\alpha$ and $M\bar{\alpha}$, respectively:
$$M^\alpha = \{e_j \in S \ / \ M\alpha(j) \neq 0\} \text{ and } M^{1-\alpha} = \{e_j \in S \ / \ M\bar{\alpha}(j) \neq 0\}.$$

- the ordered pair $(M_i^\alpha, M_i^{1-\alpha})$ of the decomposition $d_n S$ will be determined by ordered pair of $i$-th rows of the matrices $sM\alpha$ and $sM\bar{\alpha}$.

The rows and columns of matrices $sM\alpha$ and $sM\bar{\alpha}$ will be represented by the arrays $I$ and $J$ using the following notations for $i \in \{1, \ldots, n\}$ and $j \in \{1, \ldots, m\}$:

$I_i^\alpha$ is the $i$-th row of the matrix $sM\alpha$,

$J_j^\alpha$ is the $j$-th column of the matrix $sM\alpha$,

$I_i^{\bar{\alpha}}$ is the $i$-th row of the matrix $sM\bar{\alpha}$,

$J_j^{\bar{\alpha}}$ is the $j$-th row of the matrix $sM\bar{\alpha}$.

It is easy to prove that an ordered pair of $(0,1)$-matrices $(sM\alpha, sM\bar{\alpha})$ generates a special decomposition of the ordered set $S$ if the matrices satisfy the following conditions:

- $\forall i \ \forall j \ (sM\alpha(i, j) = 0) \lor (sM\bar{\alpha}(i, j) = 0)$,

- $\forall i \ [(\exists j_1 \ sM\alpha(i, j_1) \neq 0) \lor (\exists j_2 \ sM\bar{\alpha}(i, j_2) \neq 0)]$,

- $\forall j \ [(\exists i_1 \ sM\alpha(i_1, j) = 1) \lor (\exists i_2 \ sM\bar{\alpha}(i_2, j) = 1)]$.

Also, if the $\alpha$-domain of the decomposition generated by the ordered pair $(sM\alpha, sM\bar{\alpha})$ is a special covering for the set $S$, then the array $M\alpha$ contains only non-zero elements.

We denote by $(i_1, \ldots, i_k)sM\alpha$ the matrix obtained by allocating rows $I_{i_1}^{\bar{\alpha}}, \ldots, I_{i_k}^{\bar{\alpha}}$ of the matrix $sM\bar{\alpha}$ instead of corresponding rows, $I_{i_1}^\alpha, \ldots, I_{i_k}^\alpha$, of the matrix $sM\alpha$.

We say that $sM\alpha \subseteq (i_1, \ldots, i_k)sM\alpha$, if for any $j \in \{1, \ldots, m\}$, the following holds:
if for some $i$, $sM\alpha(i, j) = 1$, then for some $r$, $(i_1, \ldots, i_k)sM\alpha(r, j) = 1$.

The notation $(I_j^\alpha, I_j^{\bar{\alpha}}) \in (sM\alpha, sM\bar{\alpha})$ means that $I_j^\alpha$ and $I_j^{\bar{\alpha}}$ are $j$-th rows of the matrices $sM\alpha$ and $sM\bar{\alpha}$, respectively.

Obviously, the rows, denoted by $I_j^\alpha$ and $I_j^{\bar{\alpha}}$, correspond to the $\alpha$-component and $(1-\alpha)$-component of the ordered pair $(M_j^\alpha, M_j^{1-\alpha})$, respectively.

## 4. The Complexity of the Graph Construction Procedure

We will describe the construction of a pointing graph based on an ordered pair of matrices $(sM\alpha, sM\bar{\alpha})$, which corresponds to some special decomposition. We also continue to redefine some basic concepts using these matrices.

_Definition._ 4.1. For any $\alpha \in \{0,1\}$ and $i \in \{1, \ldots, n\}$, the row $I_i^\alpha$ of the matrix $sM\alpha$ is called $M^\alpha$-single with respect to the columns $J_{j_1}^\alpha, \ldots, J_{j_l}^\alpha$ of the matrix $sM\alpha$, if

- $(sM\alpha(i, j_1) = 1) \wedge \ldots \wedge (sM\alpha(i, j_l) = 1)$,
- $\forall (k \neq i) \, ((sM\alpha(k, j_1) = 0) \wedge \ldots \wedge (sM\alpha(k, j_l) = 0))$.

4.1.1. For any $i \in \{1, \ldots, n\}$, if $I_i^\alpha$ is the $i$th row of the $sM\alpha$ matrix and $I_i^{\bar{\alpha}}$ is the $i$th row of the $sM\bar{\alpha}$ matrix, then the permutation of the components of the ordered pair $(I_i^\alpha, I_i^{\bar{\alpha}})$ is called a replacement step or, briefly, a step.

4.1.2. We say that the replacement step $(I_i^\alpha, I_i^{\bar{\alpha}})$ leads to the replacement step $(I_k^\alpha, I_k^{\bar{\alpha}})$, where $i, k \in \{1, \ldots, n\}$, if holds the following:
- $I_i^\alpha$ is an $M^\alpha$-single row with respect to some column $J_j^\alpha$ of the matrix $sM\alpha$,
- $sM\bar{\alpha}(k, j) = 1$ $(j \in \{1, \ldots, m\})$.

4.1.3. The replacement step $(I_i^\alpha, I_i^{\bar{\alpha}})$ is called an obligatory step associated with the column $J_r^{\bar{\alpha}}$ of the matrix $sM\bar{\alpha}$ $(r \in \{1, \ldots, m\})$, if the following holds:
- $(M\alpha(r) = 0) \wedge (sM\bar{\alpha}(i, r) = 1)$,
- $I_i^{\bar{\alpha}}$ is an $M^{1-\alpha}$-single row with respect to the column $J_r^{\bar{\alpha}}$ of the matrix $sM\bar{\alpha}$.

4.1.4. The replacement step $(I_i^\alpha, I_i^{\bar{\alpha}})$ is called a possible step associated with the column $J_r^{\bar{\alpha}}$ of the matrix $sM\bar{\alpha}$ (r $\in \{1, \ldots, m\}$), if there are $\{i_1, \ldots, i_k\} \subseteq \{1, \ldots, n\}$ such that
$$(M\alpha(r) = 0) \wedge (sM\bar{\alpha}(i, r) = 1) \wedge (sM\bar{\alpha}(i_1, r) = 1) \wedge \ldots \wedge (sM\bar{\alpha}(i_k, r) = 1).$$

4.1.5. The replacement step $(I_i^\alpha, I_i^{\bar{\alpha}})$ is called a final step, if the row $I_i^\alpha$ is an immediate $M^\alpha$-replaceable row.

4.1.6. The replacement step $(I_i^\alpha, I_i^{\bar{\alpha}})$ is called a useless step, if:
- $I_i^\alpha$ is an $M^\alpha$-single row with respect to some $j$th column $J_j^\alpha$ of the matrix $sM\alpha$,
- $M\bar{\alpha}(j) = 0$ $(j \in \{1, \ldots, m\})$.

Obviously, if the matrix $sM\alpha$ does not contain columns with only zero elements, then the corresponding $\alpha$-domain is an $M^\alpha$-covering for the set $S$. So, it makes no sense to construct a graph.

If $J_{q_1}^\alpha, \ldots, J_{q_k}^\alpha$ are columns of the matrix $sM\alpha$ with only zero elements, and $I_{j_1}^{\bar{\alpha}}, \ldots, I_{j_p}^{\bar{\alpha}}$ are all rows of the matrix $sM\bar{\alpha}$ each of which contains 1s in some of the columns $J_{i_1}^{\bar{\alpha}}, \ldots, J_{i_k}^{\bar{\alpha}}$, then the set of ordered pairs denoted by $R_M$,
$$R_M = \{(I_{j_1}^\alpha, I_{j_1}^{\bar{\alpha}}), \ldots, (I_{j_p}^\alpha, I_{j_p}^{\bar{\alpha}})\},$$
will be called a set of main replacement steps for the ordered pair of matrices $(sM\alpha, sM\bar{\alpha})$.

<u>*Definition*</u> 4.2. If for some $\alpha \in \{0,1\}$, the set $R_M$,

$$R_M = \{(I_{j_1}^\alpha, I_{j_1}^{\bar\alpha}), \ldots, (I_{j_p}^\alpha, I_{j_p}^{\bar\alpha})\},$$

is a set of main replacement steps for the ordered pair of matrices $(sM\alpha, sM\bar\alpha)$, then the set $R$,

$$R = \{(I_{i_1}^\alpha, I_{i_1}^{\bar\alpha}), \ldots, (I_{i_l}^\alpha, I_{i_l}^{\bar\alpha})\},$$

will be called an $M^\alpha$-replaceability procedure based on the ordered pair $(sM\alpha, sM\bar\alpha)$, if

- $R_M \subseteq R$,

- for any replacement step $(I_k^\alpha, I_k^{\bar\alpha}) \in R \setminus R_M$, there is another replacement step $(I_j^\alpha, I_j^{\bar\alpha}) \in R$, such that $(I_j^\alpha, I_j^{\bar\alpha})$ leads to $(I_k^\alpha, I_k^{\bar\alpha})$,

- if some replacement step, $(I_j^\alpha, I_j^{\bar\alpha}) \in R$, leads to another replacement step $(I_k^\alpha, I_k^{\bar\alpha})$, then $(I_k^\alpha, I_k^{\bar\alpha}) \in R$.

If an ordered pair of matrices $(sM\alpha, sM\bar\alpha))$ corresponds to some special decomposition, then to construct the graph corresponding to the replaceability procedure based on this pair of matrices, we do the following:

Similar to the description of graph construction procedure in [8], we consider the ordered pairs of rows included in the replaceability procedure as the vertices of the graph

We denote by $v_i$ the vertex corresponding to the step $(I_i^\alpha, I_i^{\bar\alpha})$.

*If the replacement step $(I_i^\alpha, I_i^{\bar\alpha})$ is associated with the column $J_r^{\bar\alpha}$ of the matrix $sM\bar\alpha$, and the vertex $v_i$ correspond to the ordered pair $(I_i^\alpha, I_i^{\bar\alpha})$, then we say that the vertex $v_i$ is associated with the column $J_r^{\bar\alpha}$ of the matrix $sM\bar\alpha$.*

The vertices $v_{j_1}, \ldots, v_{j_p}$ corresponding to the main replacement steps

$$R_M = \{(I_{j_1}^\alpha, I_{j_1}^{\bar\alpha}), \ldots, (I_{j_p}^\alpha, I_{j_p}^{\bar\alpha})\},$$

will be called the main vertices of the graph.

The edges of the graph are formed as follows:

If the replacement step $(I_i^\alpha, I_i^{\bar\alpha})$ leads to the replacement step $(I_j^\alpha, I_j^{\bar\alpha})$, then we consider the ordered pair $(v_i, v_j)$ as a directed edge of the graph.

The labels of the edges are defined as follows:

Let $(v_i, v_j)$ be a directed edge of the graph.

- if for some $r \in \{1, \ldots, m\}$, $(I_j^\alpha, I_j^{\bar\alpha})$ is an obligatory step associated with the column $J_r^{\bar\alpha}$ of the matrix $sM\bar\alpha$, then $(v_i, v_j)$ will be called a conjunctive edge associated with the column $J_r^{\bar\alpha}$ of the matrix $sM\bar\alpha$. We add the label "&$r$" to such an edge and denote it by

$$(v_i, \&r, v_j).$$

- if for some $r \in \{1, \ldots, m\}$, $(I_j^\alpha, I_j^{\bar\alpha})$ is a possible step associated with the column $J_r^{\bar\alpha}$ of the matrix $sM\bar\alpha$, then $(v_i, v_j)$ will be called a disjunctive edge associated with the column $J_r^{\bar\alpha}$ of the matrix $sM\bar\alpha$. We add the label "∨$r$" to such an edge and denote it by

$$(v_i, \vee r, v_j).$$

If the replacement step $(I_j^\alpha, I_j^{\bar\alpha})$ is a final step, then the corresponding vertex $v_j$ will be called a final vertex. So, there are no edges outgoing from the vertex $v_j$.

If the replacement step $(I_j^\alpha, I_j^{\bar\alpha})$ is a useless step, then the corresponding vertex $v_j$ will be called a useless vertex.

<u>*Remark*</u> 4.3. If we consider the replacement step $(I_i^\alpha, I_i^{\bar\alpha}) \in (sM\alpha, sM\bar\alpha)$ such that $I_i^\alpha$ is an $M^\alpha$-single row and $I_i^{\bar\alpha}$ is an $M^{1-\alpha}$-single row then, according to the Proposition 2.3.1 in [8], there is no special covering for the set $S$ under the corresponding special decomposition. So, it makes no sense to continue any procedure.

We will represent the graph in the form of specific matrices. In parallel with the formation of these matrices, we also define and form some arrays in such a way as to have a quick access to all the necessary parameters of the graph.

The procedure for forming these arrays and matrices will be described on the basis of the ordered pair of matrices $(sM\alpha, sM\bar\alpha)$, corresponding to some special decomposition of the set $S$.

Thus, in parallel to construction procedure, we form the following arrays:

V is an array of size $n$ for the vertices of the graph:
The non-zero elements of the array V are the numbers of formed vertices of the graph in the order of their formation.

Multiplicity is an array of size $m$ whose elements are formed as follows:
If a main vertex is associated with the column $J_j^{\bar\alpha}$ of the matrix $sM\bar\alpha$ for $j \in \{1, \ldots, n\}$, then Multiplicity(j) is the number of all main vertices associated with $J_j^{\bar\alpha}$.

Multiplicity(j) = 0 for all other values of j.

Ind is an array of size $n$ formed as follows:
If some main vertex $v_i$ appears during the graph constructing procedure, then the element Ind($i$) is assigned the value 1. If an edge $(v_i, v_k)$ appears during the graph constructing procedure, then the value of Ind(k) is increased by 1.

GraphEdges is a matrix of the size $(n \times n)$ corresponding to the graph and similar to the adjacency matrix. It forms as follows:
If p is the number of directed edges outgoing from the vertex $v_i$ to the vertex $v_j$, then GraphEdges $(i, j)$ = p. GraphEdges($i, j$) = 0, otherwise.

EdgeIn is a matrix of the size $(m \times n)$ formed as follows:
For any $j \in \{1, \ldots, m\}$ and $i \in \{1, \ldots, n\}$, if an edge, denoted by $(v_r, v_i)$, is formed, then
$$\text{EdgeIn(j, } i) = r.$$
At the same time, if it is a conjunctive edge, then it is marked as a conjunctive edge, and accordingly, if it is a disjunctive edge, then it is marked as a disjunctive edge.

DisEdges is a matrix of size $(n \times m)$.

For any $i \in \{1, \ldots, n\}$ and $j \in \{1, \ldots, m\}$, the element DisEdges$(i, j)$ of the matrix DisEdges shows the number of disjunctive edges outgoing from the vertex $v_i$ and associated with the column $J_j^{\bar{\alpha}}$ of the matrix $sM\bar{\alpha}$.

In addition, if for some $i \in \{1, \ldots, n\}$, the graph does not contain the vertex $v_i$, then DisEdges$(i, j) = 0$ for any $j \in \{1, \ldots, m\}$.

We assume that at the beginning of the graph construction procedure, all those arrays and matrices have zero elements.

The construction procedure starts by determining the main replacement steps and the main vertices of the graph.

Let's note that, for convenience, sometimes, we will identify vertices of the graph with their corresponding replacement steps.

*Alg* 1: Formation of the main vertices of the graph

1.  numMainVertices = 0
2.  **For** j = 1 **to** $m$ **do**:
3.      **For** $i$ = 1 **to** $n$ **do**:
4.          **If** $sM\alpha(i, j) = 1$:                    // if $e_j$ is included in the subset $M_i^{\alpha}$
5.              $M\alpha$(j) := $M\alpha$(j) +1
6.          **Else**:
7.              **If** $sM\bar{\alpha}(i, j) = 1$:                    // if $e_j$ is included in the subset $M_i^{1-\alpha}$:
8.                  $M\bar{\alpha}$(j) := $M\bar{\alpha}$ (j) +1
9.              **EndIf**
10.         **EndFor** $i$;
11.     **EndFor** j;                    // the arrays $M\alpha$ and $M\bar{\alpha}$ are formed

12.     **For** j = 1 **to** $m$ **do**:                    // Forming the main replacement steps and main vertices
13.         **If** $M\alpha$(j) = 0:
14.             **For** r = 1 **to** $n$ **do**:                    // Looking for 1s in j-th column of $sM\bar{\alpha}$ matrix
15.                 **If** $sM\bar{\alpha}$(r, $j$) = 1:                    // $e_j$ is included in the row $M_r^{1-\alpha}$:

16.                     **If** the vertex $v_r$ is not yet formed:
17.                         Mark $v_r$ as a formed vertex
18.                         Add the vertex $v_r$ to the array V as a next non-zero element
19.                         Mark $v_r$ as a main vertex
20.                         numMainVertices := numMainVertices + 1
21.                         Ind $(v_r)$ := Ind $(v_r)$ +1
22.                     **EndIf**;
23.                     Fixing that the main vertex $v_r$ is associated with the element $e_j$
24.                     Multiplicity($e_j$) := Multiplicity($e_j$) +1

25.            **EndIf**;
26.        **EndFor** r;
27.      **EndIf**;
27.  **EndFor** j;

28.  **If** numMainVertices = 0:
29.      **print** (" $\alpha$-domain is a is a special covering for $S$")
30.      **Stop**;
31.  **Else**:
32.      Go to $Alg2$  // constructing vertices
33.  **EndIf**;

It is easy to see that this algorithm takes as input the matrices $sM\alpha$ and $sM\bar{\alpha}$ and outputs the arrays $M\alpha$ and $M\bar{\alpha}$, main vertices and the elements of the set of vertices V, that correspond to the main vertices.

We continue to construct the graph, by sequentially exploring the vertices corresponding to the elements of the array V. For each vertex $v_q \in$ V corresponding to the replacement step $(I_q^\alpha, I_q^{\bar{\alpha}})$, we do the following:

(a) if $I_q^\alpha$ is not an $M^\alpha$-single row, then there are no edges outgoing from the vertex $v_q$, which means that $v_q$ is a final vertex.

(b) if $I_q^\alpha$ is an $M^\alpha$-single row with respect to the columns $J_{j_1}^\alpha, \ldots, J_{j_k}^\alpha$, then for each $j_i \in \{j_1, \ldots, j_k\}$ we find all rows in the matrix $sM\bar{\alpha}$ containing 1s in the column $J_{j_i}^{\bar{\alpha}}$.

Let for some fixed $j_i \in \{j_1, \ldots, j_k\}$, the rows $I_{r_1}^{\bar{\alpha}}, \ldots, I_{r_l}^{\bar{\alpha}}$ contain 1 in the column $J_{j_i}^{\bar{\alpha}}$. This means that the replacement step $(I_q^\alpha, I_q^{\bar{\alpha}})$ leads to the replacement steps

$$(I_{r_1}^\alpha, I_{r_1}^{\bar{\alpha}}), \ldots, (I_{r_l}^\alpha, I_{r_l}^{\bar{\alpha}}).$$

Thus, with each $j_i \in \{j_1, \ldots, j_k\}$ the vertices $v_{r_1}, \ldots, v_{r_l}$ appear, and we form the following directed edges:

$$(v_q, v_{r_1}), \ldots, (v_q, v_{r_l}).$$

Therefore:

- we assign consecutive zero elements of array V numbers $r_1, \ldots, r_l$, if they are not already there,

- we mark the vertices $v_{r_1}, \ldots, v_{r_l}$ as already formed vertices,

- we add 1 to elements Ind $(r_1), \ldots,$ Ind $(r_l)$ of the Ind array,

- we add 1 to the values of each of the elements

$$\text{GraphEdges}(q, r_1), \ldots, \text{GraphEdges}(q, r_l)$$

of the matrix GraphEdges.

The labels for directed edges outgoing from the vertex $v_q$ are formed as follows:

Obviously, for any $j_i \in \{j_1, \ldots, j_k\}$, $M\bar{\alpha}(j_i)$ is the number of all rows of the matrix $sM\bar{\alpha}$ containing 1's in the column $J_{j_i}^{\bar{\alpha}}$. For each $i \in \{1, \ldots, k\}$ we do the following:

(b.1) if $M\bar{\alpha}(j_i) = 1$, then we have the following:

- $sM\bar{\alpha}(r_1, j_i) = 1$,
- $I_{r_1}^{\bar{\alpha}}$ is $M^{1-\alpha}$-single row with respect to the column $J_{j_i}^{\bar{\alpha}}$ in $sM\bar{\alpha}$,
- $(I_{r_1}^{\alpha}, I_{r_1}^{\bar{\alpha}})$ is the only replacement step associated with the column $J_{j_i}^{\bar{\alpha}}$.

This means that $(I_{r_1}^{\alpha}, I_{r_1}^{\bar{\alpha}})$ is an obligatory step associated with the column $J_{j_i}^{\bar{\alpha}}$ of the matrix $sM\bar{\alpha}$, and therefore, the labeled edge $(v_q, \&i_j, v_{r_1})$ is forming. So, we assign $q$ the corresponding element of the matrix EdgeIn, $\text{EdgeIn}(j_i, r_1) = q$, and mark this edge as a conjunctive edge.

(b.2) if $M\bar{\alpha}(j_i) > 1$, then the ordered pairs $(I_{r_1}^{\alpha}, I_{r_1}^{\bar{\alpha}}), \ldots, (I_{r_{l_i}}^{\alpha}, I_{r_{l_i}}^{\bar{\alpha}})$, where $l_i = M\bar{\alpha}(j_i)$, are possible replacement steps, associated with column $J_{j_i}^{\bar{\alpha}}$ and the labeled edges

$$(v_q, \vee j_i, v_{r_1}), \ldots, (v_q, \vee j_i, v_{r_{l_i}}).$$

are forming. So, we assign $q$ to the corresponding elements of the matrix $EIn$:

$$\text{EdgeIn}(j_i, r_1) = q, \ldots, \text{EdgeIn}(j_i, r_{l_j}) = q,$$

mark these edges as disjunctive edges and assign $l_j$ to the element $Vd(q, i_j)$ of the matrix $Vd$, that is $\text{DisEdges}(q, j_i) = l_i$.

(b.3) if $M\bar{\alpha}(j_i) = 0$ then $\forall s \in \{1, \ldots, n\}$ $(sM\bar{\alpha}(s, j_i) = 0)$.

This means that there is no step which follows to the step $(I_q^{\alpha}, I_q^{\bar{\alpha}})$. So, we mark $v_q$ as a useless vertex and consider other vertices in accordance with this description.

(c) the graph construction procedure will be completed if we have considered all the vertices belonging to V, and new vertices and edges are not appearing.

*Alg* 2. Constructing graph vertices

1.  newVertexIndicator = 0                          // indicates the appearance of new vertices //
2.  **While** $i \leq n$ **and** V(i) $\neq 0$:                  // $i$ shows the vertex position in the array V //
3.      **For** j = 1 **to** $m$ **do**:
4.          **If** $I_{V(i)}^{\alpha}$ is $M^{\alpha}$-single row with respect to the column $J_j^{\alpha}$:

5.              **If** $M\bar{\alpha}$(j) = 0:
6.                  Mark $v_{V(i)}$ as a useless vertex
7.                  **Update** $i$;
8.                  **Go to** Step 2;
9.              **Else**:
10.                 newVertexIndicator := 1                  // new vertices appear //

11.                 **For** r1 = 1 **to** $n$ **do**:
12.                     **If** $sM\bar{\alpha}$(r1, j) = 1:                  // if $e_j$ is included in the subset $M_{r1}^{1-\alpha}$:
13.                         GraphEdges ($v_{V(i)}$, $v_{r1}$) := GraphEdges($v_{V(i)}$, $v_{r1}$) +1
14.                         Ind ($v_{r1}$) := Ind ($v_{r1}$)+1
15.                         **If** $v_{r1}$ is not marked as an already formed vertex:
16.                             Mark $v_{r1}$ as an already formed vertex
17.                             Add $v_{r1}$ to the array V as a next non-zero element
18.                         **EndIf**;
19.                         **If** $sM\bar{\alpha}$(j) = 1:
20.                             Mark ($v_{V(i)}$, $v_{r1}$) as a conjunctive edge ($v_{V(i)}$, &j, $v_{r1}$)
21.                         **Else**:
22.                             Mark ($v_{V(i)}$, $v_{r1}$) as a disjunctive edge ($v_{V(i)}$, Vj, $v_{r1}$)
23.                             DisEdges ($v_{V(i)}$, j) := DisEdges ($v_{V(i)}$, j)+1
24.                         **EndIf**;
25.                     **EndIf**;
26.                 **EndFor** r1;

27.             **EndIf**;
28.         **EndIf**;
29.       **EndFor** j;
30.   **EndWhile** $i$;

31.   **If** newVertexIndicator = 0:
32.       **Output** ("Alpha domain corresponding to the main vertices is a special covering")
33.       **Stop**;
34.   **Else**:
35.       **Go to** $Alg3$
36.   **EndIf**;

Thus, if we are given that:

- $S = \{e_1, e_2, \ldots, e_m\}$ a is non-empty set of $m$ elements,

- the ordered pair of matrices ($sM\alpha$, $sM\bar{\alpha}$) corresponds to some special decomposition of the set $S$,

the algorithms $Alg1$ and $Alg2$ form all the vertices and edges of the graph. At the same time, the arrays $M\alpha$, $M\bar{\alpha}$, V, Ind, and the matrices GraphEdges, EdgeIn, DisEdges are formed as a result of these algorithms.

_Definition_ 4.4. The total number of non-zero elements of the matrices $sM\alpha$ and $sM\bar{\alpha}$ will be called the length of input data of the special decomposition.

Once the main vertices of the graph are formed, subsequent replacement steps produce additional vertices leading to further replacement steps.

According to graph construction procedure, we explore each vertex to find out if there are outgoing edges from it. Therefore, the following Remark is important.

<u>*Remark*</u> 4.5. (a) If during the replaceability procedure the step $(I_i^\alpha, I_i^{\overline{\alpha}})$ leads to some step $(I_k^\alpha, I_k^{\overline{\alpha}})$, then we consider $(I_k^\alpha, I_k^{\overline{\alpha}})$ to find out if it is an obligatory or a possible step, and also to find out if $I_k^\alpha$ is an $M^\alpha$-single row. If another replacement step $(I_j^\alpha, I_j^{\overline{\alpha}})$ leads to the already considered step $(I_k^\alpha, I_k^{\overline{\alpha}})$, then it makes no sense to consider it again.

Thus, although some of the replacement steps may occur more than once in the replaceability procedure, each of them is explored once.

4.5. (b) If $(I_i^\alpha, I_i^{\overline{\alpha}})$ and $(I_j^\alpha, I_j^{\overline{\alpha}})$ are different ordered pairs included in the description of the replaceability procedure, then the rows $I_i^\alpha$ and $I_j^\alpha$ cannot be $M^\alpha$-single rows with respect to the same column of the matrix $sM\alpha$.

<u>*Definition*</u> 4.6. The following operations will be called elementary:
- assigning a number to a variable or array element,
- addition and subtraction of numbers,
- comparison of two numbers.

Our goal is to find an estimate of the number of elementary operations required to complete the graph construction procedure by comparing it with the length of the input data of a special decomposition.

For some $q \in \{1, \ldots, n\}$ and $\alpha \in \{0,1\}$, we make the following notation:
$N(d_nS)$ is the length of input data of the decomposition $d_nS$,
$m_q^\alpha$ is the number of non-zero elements of the row $I_q^\alpha$,
$m_q^{\overline{\alpha}}$ is the number of non-zero elements of the row $I_q^{\overline{\alpha}}$,
$EG$ is the number of edges of the graph under construction.
It is easy to notice that
$$m_q^\alpha + m_q^{\overline{\alpha}} \leq m,$$
$$N(d_nS) = m_1^\alpha + m_1^{\overline{\alpha}} + \ldots + m_n^\alpha + m_n^{\overline{\alpha}} \leq n \times m.$$

The following lemma gives an important result for estimating the number of edges of a pointing graph. Actually, we obtain an upper bound on the number of edges in the pointing graph.

<u>*Lemma*</u> 4.7. For any pointing graph corresponding to an $M^\alpha$-replaceability procedure based on the ordered pair of matrices $(sM\alpha, sM\overline{\alpha})$ of size $(n \times m)$,
$$EG \leq (n - 1) \times m.$$

<u>*Proof.*</u> The edges of the graph appear only when an ordered pair is included in the replaceability procedure, let it be the pair $(I_j^\alpha, I_j^{\overline{\alpha}})$, such that $I_j^\alpha$ is an $M^\alpha$-single row with respect to some columns of the matrix $sM\alpha$.

Let $v_{i_1}, v_{i_2}, \ldots, v_{i_l}$ be all vertices of the graph with outgoing edges. It means that in the $M^\alpha$-replaceability procedure there are ordered pairs

$$(I_{i_1}^\alpha, I_{i_1}^{\bar{\alpha}}), (I_{i_2}^\alpha, I_{i_2}^{\bar{\alpha}}), \ldots, (I_{i_l}^\alpha, I_{i_l}^{\bar{\alpha}})$$

such that $I_{i_1}^\alpha, I_{i_2}^\alpha, \ldots, I_{i_l}^\alpha$ are $M^\alpha$-single rows. We denote by $l_{i_j}^\alpha$, where $j \in \{1, \ldots, l\}$, the number of columns in the matrix $sM\alpha$ with respect to which the row $I_{i_j}^\alpha$ is $M^\alpha$-single.

Suppose that (worst case) for each $j \in \{1, \ldots, l\}$, all rows of the matrix $sM\bar{\alpha}$, except for $I_{i_j}^{\bar{\alpha}}$, contain 1 in all $l_{i_j}^\alpha$ columns, with respect to which the row $I_{i_j}^\alpha$ is $M^\alpha$-single. In this case, the number of edges outgoing from the vertex $v_{i_j}$ is equal to $(\text{n - 1}) \times l_{i_j}^\alpha$.

Therefore,

$$EG = (n - 1) \times (l_{i_1}^\alpha + l_{i_2}^\alpha + \ldots + l_{i_l}^\alpha).$$

Since different rows cannot be $M^\alpha$-single with respect to the same column, then

$$l_{i_1}^\alpha + l_{i_2}^\alpha + \ldots + l_{i_l}^\alpha \leq m,$$

So, we obtain that

$$EG \leq (n - 1) \times m. \ \nabla$$

_Proposition_ 4.8. The number of elementary operations to determine the main vertices of the $M^\alpha$-pointing graph corresponding to an ordered pair of matrices $(sM\alpha, sM\bar{\alpha})$ does not exceed the number

$$c \times n \times m$$

for some constant $c$.

_Proof_. The algorithm $Alg\,1$ takes as an input an ordered pair $(sM\alpha, sM\bar{\alpha})$, and outputs the arrays $M\alpha$, $M\bar{\alpha}$ and the main vertices of the graph.

To find elements of the arrays $M\alpha$ and $M\bar{\alpha}$, the algorithm runs through all the elements of both matrices, compares them to the numbers 0 or 1, and sums all the 1s in each column.

Obviously, all these operations are elementary.

To determine the main vertices, the algorithm performs the following operations:

a) The algorithm finds columns with zero elements in the matrix $sM\alpha$. To do this it suffices to find the zero elements of the $M\alpha$ array, which can be found in $m$ elementary operations. Let $J_{q_1}^\alpha, \ldots, J_{q_l}^\alpha$ be the zero columns of the matrix $sM\alpha$.

b) The algorithm finds all rows of the matrix $sM\bar{\alpha}$ that contain 1 in any of the columns $J_{q_1}^{\bar{\alpha}}, \ldots, J_{q_l}^{\bar{\alpha}}$. Obviously, this can be done in $n \times l$ elementary operations.

c) The algorithm assigns corresponding values to the elements of the array V. This can be done in at most $c \times n$ elementary operations for some constant $c$.

So, the total number of elementary operations to determine the main vertices of the graph does not exceed the number $c \times n \times m$ for some constant $c$. $\nabla$

*Proposition* 4.9. The total number of elementary operations for finding all of the replacement steps with an $M^\alpha$-single $\alpha$-component, and determining all ordered pairs that these replacement steps lead to, does not exceed the number

$$c \times n \times m$$

for any replaceability procedure.

*Proof*. Recall that the new edges and new vertices of the graph appear only if the replaceability procedure includes an ordered pair $(I_i^\alpha, I_i^{\bar\alpha})$, such that $I_i^\alpha$ is an $M^\alpha$-single row with respect to some columns of the matrix $sM\alpha$. Recall also, that each element of the array V corresponds to an ordered pair included in the replaceability procedure.

The algorithm $Alg\,2$ examines the array V to find out if there are elements such that the $\alpha$-components of the ordered pairs corresponding to these elements are $M^\alpha$-single subsets.

If the pair $(I_{V(i)}^\alpha, I_{V(i)}^{\bar\alpha})$ corresponds to the element $V(i)$, then the algorithm searches for 1s in the row $I_{V(i)}^\alpha$ such that each of them is the only one in the column containing it.

Obviously, if $sM\alpha(V(i), j) = 1$ and $M\alpha(j) = 1$, then $I_{V(i)}^\alpha$ is an $M^\alpha$-single row with respect to the column $J_j^\alpha$ of $sM\alpha$.

If these conditions are satisfied, then the algorithm proceeds to find 1s in the column $J_j^{\bar\alpha}$ of the matrix $sM\bar\alpha$.

Thus, for any element $V(i)$ the algorithm performs the following operations:

- finds 1 in the row $I_{V(i)}^\alpha$ in no more than $m$ elementary operations,

- if this 1 is in the column $J_j^\alpha$, that is, $M\alpha(V(i), j) = 1$ and at the same time $M\alpha(j) = 1$, then the algorithm proceeds to look for 1s in the column $J_{p_j}^{\bar\alpha}$ of the matrix $sM\bar\alpha$. These 1s will determine the ordered pairs where $(I_{V(i)}^\alpha, I_{V(i)}^{\bar\alpha})$ leads.

Obviously, no more than $n$ elementary operations will be required.

Note, that the number of $M^\alpha$-single rows in the matrix $sM\alpha$ cannot exceed the number $m$, since different rows in $sM\alpha$ cannot be $M^\alpha$-single with respect to the same column.

This means that the number of columns in the matrix $sM\alpha$, in which only one element is equal to 1, does not exceed the number $m$. For any of these columns, the algorithm performs $n$ elementary operations in order to find 1s in the corresponding column of the matrix $sM\bar\alpha$. These 1s determine the ordered pairs that the previous replacement step leads to.

That is, if

- the ordered pair $(I_{V(i)}^\alpha, I_{V(i)}^{\bar\alpha})$ corresponds to the element $V(i)$,

- the row $I_{V(i)}^\alpha$ is $M^\alpha$-single with respect to the columns $J_k^\alpha$,

- $(sM\bar\alpha(V(i), k_1) = 1) \wedge \ldots \wedge (sM\bar\alpha(V(i), k_p)=1)$ for some $\{k_1, \ldots, k_p\} \subseteq \{1, \ldots, n\}$,

then the ordered pair $(I_{V(i)}^\alpha, I_{V(i)}^{\bar\alpha})$ leads to the ordered pairs

$$(I_{k_1}^\alpha, I_{k_1}^{\bar\alpha}), \ldots, (I_{k_p}^\alpha, I_{k_p}^{\bar\alpha}).$$

So, in the graph are formed the edges

$$(v_{V(i)}, v_{k_1}), \ldots, (v_{V(i)}, v_{k_p}),$$

which will be represented by means of the matrix GraphEdges and the matrix EdgeIn.

Thus, the total number of elementary operations to find all replacement steps with an $M^\alpha$-single $\alpha$-component, and to determine all ordered pairs that these steps lead to, does not exceed the number $c \times n \times m$ for any replaceability procedure. $\nabla$

*Corollary* 4.9.1. The number of elementary operations required to find all vertices and form all edges with their labels of any pointing graph does not exceed the number

$$c \times n \times m$$

for some constant $c$.

*Proof.* According to Propositions 4.8 and 4.9, all vertices and edges of the graph are formed in at most $c \times n \times m$ elementary operations for some constant $c$.

When forming an edge, we determine if it is a conjunctive edge or a disjunctive edge.

Let the edge $(v_i, v_j)$ be formed. This means that the row $I_i^\alpha$ is an $M^\alpha$-single with respect to some $p$-th column $J_p^\alpha$ of the matrix $sM\alpha$ and $sM\bar{\alpha}(j, p) = 1$. To determine if this edge is conjunctive or disjunctive, we check the corresponding value of the array $M\bar{\alpha}$. If $M\bar{\alpha}(p) = 1$, then $(v_i, v_j)$ is a conjunctive edge, and if $M\bar{\alpha}(p) > 1$, then it is a disjunctive edge.

Taking into account also the fact that according to Lemma 3.6, the number of graph edges does not exceed the number $(n - 1) \times m$, it is obvious, that these additional elementary actions do not affect the order of the resulting estimate for finding edges and vertices. $\nabla$

*Corollary* 4.9.2. The number of elementary operations required to form the arrays V, Multiplicity, Ind and the matrices GraphEdges, EdgeIn, DisEdges does not exceed the number

$$c \times n \times m$$

for some constant $c$.

*Proof.* During the constructing procedure of the graph, we deal with these arrays and matrices only when we are looking for main vertices and when new edges are formed. In both procedures, we perform at most one elementary operation with each of them. Therefore, according to the Propositions 4.8 and 4.9, the number of elementary actions necessary to form all these arrays does not exceed the number

$$c \times n \times m$$

for the certain constant $c$. $\nabla$

*Remark* 4.10. Obviously, total number of non-zero elements of the matrices $sM\alpha$ and $sM\bar{\alpha}$ coincides with the total number of the elements included in all subsets of the decomposition corresponding to the ordered pair of matrices $(sM\alpha, sM\bar{\alpha})$.

Therefore, combining the results of Propositions 3.7 and 3.8, we can assert:

In order to construct a pointing graph based on some special decomposition of a given set and form the arrays Multiplicity, V, Ind and the matrices GraphEdges, EdgeIn, DisEdges, we need at most $c \times n \times m$ elementary operations for some constant $c$.

## 5. Complexity of the Cleaning Procedure

Obviously, the replaceability procedure will be completed if each path in the corresponding pointing graph, started from any of the main vertices, reaches some final vertex.

If useless vertices are formed in the graph, the corresponding sequence of replacement steps will not reach the final step, so the replaceability procedure will not be completed.

Therefore, if the $\alpha$-domain obtained after performing the replacement steps corresponding to the vertices of the constructed graph is not a special covering for the set $S$, we first look for useless vertices in the graph in order to eliminate them.

Recall that the $\alpha$-domain of the decomposition is an $M^{\alpha}$-covering for the set if

$$\forall\, j \in \{1, \ldots, m\}\, (M\alpha(j) > 0).$$

Let's describe the procedure of eliminating a useless vertex in the graph based on the pair of matrices $(sM\alpha,\, sM\bar{\alpha})$. To do this, we will use arrays and matrices formed during the construction of the graph.

### 5.1. *Removal Procedure in The Graph*

The procedure of successive removal of certain vertices and edges in the $M^{\alpha}$-pointing graph described below will be called the removal procedure.

The procedure starts by removing a specific vertex $v$ of the graph and consists of the following steps:

5.1.1. We remove the vertex $v$ and all outgoing edges of this vertex.

If, as a result of the removal of these edges, vertices with zero indegree appear, we mark them so that we can remove them later.

We will call these vertices the generations of the already removed vertex.

5.1.2. We sequentially examine the edges entering the vertex $v$.

Let all edges outgoing from some vertex $v_j$ and entering vertex $v$ be associated with the columns $J^{\alpha}_{r_1}, \ldots, J^{\alpha}_{r_k}$, respectively.

If they are all disjunctive edges, denote them by

$$(v_j, \vee r_1, v), \ldots, (v_j, \vee r_k, v),$$

such that $\mathrm{DisEdges}(j, r) > 1$ for any $r \in \{r_1, \ldots, r_k\}$, then we remove all these edges and consider other incoming edges to the vertex $v$, if there are any.

If for some $r \in \{r_1, \ldots, r_k\}$, $(v_j, \vee r, v)$ is a single-disjunctive edge, $(\mathrm{DisEdges}(j, r) = 1$ because the other disjunctive edges associated with the column $J^{\alpha}_r$ have already been removed), or it is a conjunctive edge $(v_j, \&r, v)$ then:

- we remove all edges outgoing from the vertex $v_j$ and entering the vertex $v$,

- we mark the vertex $v_j$ so that we can remove it later, and move on to considering other incoming edges to the vertex $v$, if there are any.

The marked vertices will be called the ancestors of the already removed vertex.

5.1.3. to continue the procedure, we consider the marked vertices sequentially:

- if the vertex under consideration is marked as a generation, point 5.1.1) applies to it.

- if the vertex under consideration is marked as an ancestor, we remove this vertex, and apply points 5.1.1) and 5.1.2) to it.

The procedure ends if there are no more marked vertices.

In other words, if the edges
$$(v_{j_1}, v_i), \ldots, (v_{j_r}, v_i) \text{ and } (v_i, v_{k_1}), \ldots, (v_i, v_{k_s})$$
are included in the graph and we remove the vertex $v_i$, then all edges
$$(v_{j_1}, v_i), \ldots, (v_{j_r}, v_i) \text{ and } (v_i, v_{k_1}), \ldots, (v_i, v_{k_s})$$
are removed.

The vertex $v_j \in \{v_{j_1}, \ldots, v_{j_r}\}$ is removed only if $(v_j, v_i)$ is a conjunctive edge, or $(v_j, v_i)$ is a single-disjunctive edge associated with some column.

- the vertex $v_k \in \{v_{k_1}, \ldots, v_{k_s}\}$ is removed only if, as a result of removing the edges
$$(v_i, v_{k_1}), \ldots, (v_i, v_{k_s}),$$
the indegree of the vertex $v_k$ becomes zero.

*Recall that the vertex $v$ is called removable if, as a result of applying the removal procedure to it, for any $j \in \{1, \ldots, m\}$, some main vertex associated with the column $J_j^\alpha$ is not removed* [8].

### 5.2. *The Cleaning Procedure of The Graph.*

The procedure described below for eliminating the useless vertices in the $M^\alpha$-pointing graph will be called a cleaning procedure of the graph.

As it is described in [8], the idea of cleaning a graph is to remove useless vertices and possibly some other vertices and edges of the graph without violating the stability of the set of elements not included in the $\alpha$-domain of the decomposition under consideration.

To do it, we use the removal procedure. If a useless vertex is found in the graph, we apply the Removal Procedure to it. If it turns out that it is a removable vertex, we conclude that the current useless vertex has been eliminated, and proceed to eliminate other useless vertices in the resulting graph, if there are any.

If we find a useless vertex that is not removable, we conclude that it is impossible to eliminate useless vertices in the graph, and the cleaning procedure is terminated. We also conclude that there is no special covering for a given set under the given decomposition.

*The graph $G(e_{q_1}, \ldots, e_{q_l})$ is called clean if it does not contain useless vertices and for any element $e \in \{e_{q_1}, \ldots, e_{q_l}\}$ it contains some main vertex associated with $e$.*

Let's first describe the $Alg3$ algorithm, which executes the removal procedure started by removing a certain vertex. This algorithm will be used as a subroutine in the cleaning and compatibility procedures.

During the procedure, for convenience, the algorithm generates and uses the following one-dimensional arrays of the size $n$:

Anc is an array of ancestor vertices of size $n$. A vertex is added to this array if it is an ancestor of a removed vertex.

Gen is an array of generation vertices of size $n$. A vertex is added to this array if it is a generation of a removed vertex.

Rem is an array of size $n$ whose elements indicate whether a given vertex is removable.

RemovedVertices is an array of size $n$. The non-zero elements of this array indicate the vertices that are removed during the removal procedure.

It is assumed that all initial elements of these arrays are zero.

*RP*. (Removal Procedure).

The algorithm is a subroutine which removes vertices from the graph according to the removal procedure.

As input, the algorithm receives the index of the vertex to which the removal procedure is applied. This index is assigned to the first Anc(1) element of the *Anc* array.

The algorithm uses the following counters:

formAnc is a counter for adding vertices to the array Anc,

formGen is a counter for adding vertices to the array Gen,

counterAnc is a counter for considering the elements of the array Anc,

counterGen is a counter for considering the elements of the array Gen.

The removal procedure starts from the vertex with number Anc(1).

RP. (Removal Procedure)

1. **Initialize variables** formAnc, formGen, counterAnc, counterGen
2. **While** counterAnc $\leq n$ **and** Anc(counterAnc) $\neq 0$:
3.    p := Anc(counterAnc)
4.    Mark $v_p$ as a removed vertex
5.    Add $v_p$ to the array RemovedVertices as a next non-zero element

6.    **If** $v_p$ is marked as a main vertex:
7.       el := the column of the matrix $sM\alpha$ that $v_p$ is associated with
8.       **If** Multiplicity (el) = 1:        // if $v_p$ is the only main vertex associated with element eL
9.          Rem (Anc(1)) := -1                    // $v_p$ is not removable vertex
10.          **Return**
11.       **Else:**
12.          Multiplicity (el) := Multiplicity (el) -1
13.       **EndIf;**
14.    **Else:**

15.       **For** any vertex $v_r$ of the graph **until** all vertices are considered **do**:

16.        **If** the edge $(v_r, v_p)$ is included in the graph:

17.         Find columns of $sM\bar{\alpha}$ that the edges going from $v_r$ to $v_p$ are associated with

18.        **If** all edges from $v_r$ to $v_p$ are non-single disjunctive edges:

19.         GraphEdges (r, p) := 0              //Remove all edges from $v_r$ to $v_p$

20.         DisEdges (r, $i$) := DisEdges (r, $i$) − 1

21.        **Else:**         //if there is a conjunctive edge or a single disjunctive edge //

22.         **if** the vertex $v_r$ is not marked as an ancestor:

23.          Assign r to the next non-zero element of the array Anc

24.          GraphEdges (r, p) := 0

25.         **EndIf**;

26.        **EndIf**;

27.       **EndIf**;

28.      **EndFor** vertices;

29.      **For** any vertex $v_s$ of the graph **until** all vertices are considered **do**:

30.        **If** the edge $(v_p, v_s)$ is included in the graph:

31.         Ind(s) := Ind(s) - GraphEdges (p, s)

32.         GraphEdges (p, s) := 0          // the edges from $v_p$ to $v_s$ are removed

33.        **If** Ind(s) = 0:

34.         Assign s to the next non-zero element of the array Gen

35.        **Else:**

36.        **Endif**;

37.       **EndIf**;

38.      **EndFor** vertices;

39.    **EndWhile** counterAnc;

40.    **While** counterGen $\leq n$ **and** Gen(counterGen) $\neq 0$:

41.     q := Gen(counterGen)

42.     **If** the vertex $v_q$ is not marked as a removed vertex:

43.      Mark $v_q$ as a removed vertex

44.      Add $v_q$ to the array RemovedVertices as a next non-zero element

45.      **For** any vertex $v_s$ of the graph **until** all vertices are considered **do**:

46.        **If** the edge $(v_q, v_s)$ is included in the graph:

47.         Ind (s) := Ind (s) - GraphEdges (q, s)

48.         GraphEdges (q, s) := 0

49.        **If** Ind (s) = 0:

50.         Assign s to the next non-zero element of the array Gen

51.        **EndIf**;

52.       **Endif**;

53.      **EndFor** vertices;

54.    **EndIf**;

55. **EndWhile** counterGen;

56. Rem(Anc(1)) := 1

57. **Return**

_Proposition_ 5.3. The number of elementary operations when applying the removal procedure to any vertex of the graph does not exceed the number

$$c \times n \times (n + m)$$

for some constant $c$.

_Proof_. Assume that the removal procedure is applied to the vertex $v_i$.

The algorithm $Alg3$ does just as it is described in 5.1, that is:

- adds the number $i$ to the array $Anc$ and searches for vertices with outgoing edges entering the vertex $v_i$,

- if all edges coming out from some vertex $v_s$ to $v_i$ are disjunctive, but not single-disjunctive, then they are all removed leaving the vertex $v_s$ unchangeable.

- if a conjunctive edge or a single-disjunctive edge enters the vertex $v_i$, leaving some vertex $v_s$, then all edges leaving $v_s$ and entering the vertex $v_j$ are removed. The number $s$ is added to the array $Anc$ for subsequent removal of vertex $v_s$.

To find all these vertices, the $Alg3$ algorithm looks for non-zero elements of the $i$-th row of the EdgeIn matrix. These elements show the edges incoming the vertex $v_i$. The algorithm also finds out whether the edges that enter the vertex $v_p$ are conjunctive or disjunctive using marks of these edges, or using the matrix DisEdges.

If for some $j \in \{1, \ldots, m\}$, EdgeIn$(j, i) = s \neq 0$ and DisEdges$(s, j) > 1$, then there is a disjunctive edge $(v_s, \vee j, v_i)$ which is not single-disjunctive,

If for some $j \in \{1, \ldots, m\}$, EdgeIn$(j, i) = s \neq 0$ and DisEdges$(s, i) \leq 1$, then it is a conjunctive edge $(v_s, \& j, v_i)$, or a single-disjunctive edge $(v_s, \vee r, v_i)$.

Obviously, to do this the algorithm needs no more than $c \times m$ elementary operations for some constant $c$.

After removing the incoming edges of the vertex $v_j$, the algorithm removes the edges outgoing from it.

If as a result of the removal of all outgoing edges from $v_i$ to some vertex $v_s$, the indegree of $v_s$ becomes zero, this vertex is added to the array Gen.

To find the vertices where the edges outgoing from $v_i$ enter, the algorithm considers the $i$-th row of the matrix GraphEdges.

If GraphEdges$(i, s) > 0$ for some $s \in \{1, \ldots, n\}$, then the graph contains edges outgoing from the vertex $v_i$ and entering the vertex $v_s$. These edges are removed by assigning zero to the element GraphEdges$(i, s)$ of the matrix GraphEdges.

Obviously, to do this the algorithm needs no more than $c \times n$ elementary operations for some constant $c$.

Thus, the number of elementary operations to find and remove incoming and outgoing edges of the vertex $v_i$ does not exceed the number $c \times (m + n)$ for some constant $c$.

The algorithm continues the same procedure for the vertices corresponding to other non-zero elements of the arrays Anc.

In addition, the algorithm performs a similar procedure for vertices corresponding to non-zero elements of the Gen array, since these arrays contain only the indices of the vertices to be removed.

The arrays Anc and Gen can intersect, that is, the same vertex can be an ancestor of one removed vertex and a generation of another removed vertex at the same time. Therefore, before the applying the Removal Procedure to the vertex included in the Anc array or in the Gen array, the algorithm checks whether this vertex has already been removed during another procedure.

Thus, with the removal of any vertex the algorithm performs $c \times (m + n)$ elementary operations for some constant $c$, which means that during the procedure, the number of elementary operations depends only on the number of removed vertices.

If $l$ is the total number of removed vertices when the removal procedure is applied to the vertex $v_i$, then the total number of elementary operations to complete this removal procedure does not exceed the number $c \times l \times (n + m)$ for some constant $c$.

Since $l \leq n$, then the number of elementary operations to complete the removal procedure applied to any vertex of the graph does not exceed the number

$$c \times n \times (n + m)$$

for some constant $c$. $\nabla$

Essentially, the cleaning procedure is the application of a removal procedure to the useless vertices of the graph.

Based on removal procedure, it is easy to describe a procedure that eliminates all useless vertices in any graph, if they are removable.

We will proceed as follows:

We are looking for a useless vertex and, if we find one, we apply the removal procedure to this vertex.

If the removal procedure is completed and this vertex is eliminated, we continue the same procedure for another useless vertex until they are all eliminated.

If we find a useless vertex that cannot be eliminated, then we conclude that the graph cannot be cleaned.

We terminate all procedures, since obviously there cannot be a special covering under given special decomposition.

Note that we can find useless vertices by their mark.

*Alg*3. *Cleaning Procedure in the Graph*

1.  **For** $i := 1$ **to** $n$ **do:**
2.      **If** ($v_i$ is a useless vertex) **and** ($v_i$ has not been removed yet):
3.          Anc (1) := $i$
4.          **Call** *RP*;                                    // Apply the *Removal Procedure* to Anc (1)

5.          **If** Rem (Anc (1)) = -1:
6.              Output ('the useless vertex $v_i$ is not eliminated.')
7.              **Stop;**
8.          **EndIf;**

9.      **EndIf;**
10. **EndFor;**

Let's now estimate the complexity of the graph cleaning procedure.

*Theorem* 5.4. The total number of elementary operations for eliminating all useless vertices in the graph does not exceed the number

$$c \times n(n + m)$$

for some constant $c$.

*Proof*. Assume that $v_{i_1}, \ldots, v_{i_q}$ are useless vertices in the graph.

According to Lemma 5.7.1 in [8], if the pointing graph includes different useless vertices, then the result of the cleaning procedure does not depend on the order in which the removal procedure is successively applied to each of them.

The useless vertices are marked during the graph constructing procedure.

The number of useless vertices in any pointing graph can only be increased as a result of graph extension procedure, and this number is decreased when applying the removal procedure to the useless vertex which is removable. According to the description of the removal procedure, removed vertices do not restored.

Note that during the removal procedure applied to any useless vertex, also some other useless vertices may be removed. Therefore, before applying the Removal Procedure to another vertex, the algorithm checks whether this vertex has already been removed during another removal procedure.

Also, applying the removal procedure to any useless vertex, certain vertices and edges of the graph are removed. The indices of vertices to be removed are collected in the Anc array.

With the removal of any vertex $v_q$, the index of another vertex $v_p$ will be added to the array Anc only, if a conjunctive edge or a single-disjunctive edge enters the vertex $v_q$, leaving the vertex $v_p$. In this case, all edges leaving $v_p$ and entering the vertex $v_q$ are removed.

The number p is added to the Anc array for subsequent removal of vertex $v_p$.

If the vertex $v_p$ is also an ancestor of another removed vertex, then, according to the description of the algorithm for removal procedure, $v_p$ will not be added to the Anc array again. So, the operations will not be repeated.

Also, if a vertex is removed as an ancestor during some removal procedure, then it cannot be included in the array $Anc$ during another removal procedure, since it is no longer in the graph.

The index of some vertex $v_s$ will be added to the array Gen when the vertex $v_q$ is removed, only if the indegree of the vertex $v_s$ becomes equal to zero after the removal of $v_q$.

If the index of the vertex $v_s$ is added to the array Gen, then the algorithm performs the same procedure for this vertex when it is reached.

It is important to note that a vertex cannot be a generation vertex for more than one vertex at the same time. If several edges enter the vertex $v$ from different vertices, then with removal of each of these vertices only the following operations are performed with respect to the vertex $v$:

- the indegree of the vertex $v$ is decreased by the number of edges that enter $v$ from the removed vertex,

- the updated indegree of $v$ is compared to zero.

The index of the vertex $v$ will be added to the Gen array only if all vertices with outgoing edges entering vertex $v$ are removed.

This means that applying the removal procedure to different useless vertices, the same aforementioned operations with the same vertices will not be repeated in different procedures.

Let $n_j$ be the total number of removed vertices when the removal procedure is applied to the useless vertex $v_{i_j}$.

Using the same reasoning as in the proof of Proposition 5.3, we can assert that the number of elementary operations to complete the removal procedure applied to the useless vertex $v_{i_j}$ does not exceed the number $c \times n_j \times (n + m)$ for some constant $c$.

The total number of removed vertices during the cleaning procedure does not exceed the number of vertices of the graph. That is,

$$n_1 + \ldots + n_q \leq n.$$

Therefore, then the number of elementary operations for cleaning the pointing graph does not exceed the number

$$c \times n \times (n + m)$$

for some constant $c$. $\nabla$

According to the Theorem 6.3 in [8], if the graph does not contain useless vertices or all they are eliminated, we look for incompatible sets of vertices in the graph in order to eliminate them.

# 6. Incompatible Sets of Vertices in a Graph

We are given the pair of matrices $(sM\alpha, sM\bar{\alpha})$ of size $(n \times m)$ corresponding to some special decomposition and $\{q_1, \ldots, q_l\}$ are the indices of zero columns of the matrix $sM\alpha$.

Let $G(q_1, \ldots, q_l)$ be a graph corresponding to the replaceability procedure based on the ordered pair of these matrices, and $CG(q_1, \ldots, q_l)$ be a clean graph corresponding to the graph $G(q_1, \ldots, q_l)$.

_Definition_ 6.1. Let for some $\alpha \in \{0,1\}$
$$R = \{(I_{i_1}^{\alpha}, I_{i_1}^{\bar{\alpha}}), \ldots, (I_{i_l}^{\alpha}, I_{i_l}^{\bar{\alpha}})\}$$
be the set of replacement steps corresponding to the vertices of the graph $CG$.

We will say that the set of replacement steps
$$\{(I_{r_1}^{\alpha}, I_{r_1}^{\bar{\alpha}}), \ldots, (I_{r_k}^{\alpha}, I_{r_k}^{\bar{\alpha}})\} \subseteq R$$
is an incompatible subset in the set $R$ with respect to some column $J_q^{\alpha}$ of the matrix $sM\alpha$ if the following holds:

- $(k > 1) \wedge [\forall i \in \{r_1, \ldots, r_k\} \, (sM\alpha(i, q) = 1)]$,
- $\forall i \in (\{1, \ldots, m\} \setminus \{r_1, \ldots, r_k\}) \, (sM\alpha(i, q) = 0)$,
- $\forall i \in (\{i_1, \ldots, i_l\} \setminus \{r_1, \ldots, r_k\}) \, (sM\bar{\alpha}(i, q) = 0)$.

Respectively, the corresponding set of vertices of the graph, $\{v_{r_1}, \ldots, v_{r_k}\}$, will be called incompatible with respect to the column $J_q^{\alpha}$ of the matrix $sM\alpha$.

Let $\{v_{i_1}, \ldots, v_{i_l}\}$ be the set of vertices of a clean graph $CG$.

To find out if the graph $CG$ contains an incompatible set of vertices, we do the following

- form the special decomposition corresponding to the vertices of the clean graph and represent it with a new pair of $(0, 1)$ matrices,
- find the elements of the arrays $M\alpha$ and $M\bar{\alpha}$ under the formed decomposition,
- look for zero elements in the array $M\alpha$.

Since $\{v_{i_1}, \ldots, v_{i_l}\}$ is the set of vertices of the clean graph, it is easy to see that the new ordered pair of $(0, 1)$ matrices, denoted by $(sM\beta, sM\bar{\beta})$, which will represent the special decomposition corresponding to the clean graph can be formed as follows:

For any $i \in (\{1, \ldots, n\}$ and $j \in (\{1, \ldots, m\}$,

$$sM\beta(i, j) = \begin{cases} sM\alpha(i, j), & \text{if } i \notin \{i_1, \ldots, i_l\} \\ sM\bar{\alpha}(i, j), & \text{if } i \in \{i_1, \ldots, i_l\} \end{cases}, \quad sM\bar{\beta}(i, j) = \begin{cases} sM\bar{\alpha}(i, j), & \text{if } i \notin \{i_1, \ldots, i_l\} \\ sM\alpha(i, j), & \text{if } i \in \{i_1, \ldots, i_l\} \end{cases}.$$

Accordingly, the elements of the arrays $M\alpha$ and $M\bar{\alpha}$ will be updated on the base of the new formed pair of matrices $(sM\beta, sM\bar{\beta})$.

Recall that by Theorem 6.3 in [8], if the $\alpha$-domain of the decomposition corresponding to the vertices of the clean graph is not an $M^\alpha$-covering for the set $S$, then the clean graph contains incompatible sets of vertices.

This means that if the array $M\alpha$ contains zero elements, then the graph contains incompatible sets of vertices.

Let $M\alpha(j) = 0$ for some $j \in \{1, \ldots, m\}$.

Since $M\alpha(j) = 0$ after the permutation of ordered pairs corresponding to the vertices of the clean graph, this means that all 1s of the column $J_j^\alpha$ of the matrix $sM\alpha$ are moved to the $\bar\alpha$-domain. On the other hand, none of the 1s included in the column $J_j^{\bar\alpha}$ is moved to the $\alpha$-domain. Therefore, to find incompatible set of vertices, we look for 1s in the column $J_j^\alpha$ of the matrix $sM\alpha$.

Note that, in this case the column $J_j^\alpha$ of the matrix $sM\alpha$ definitely contains 1s, since the zero columns of the matrix $sM\alpha$ generate the main vertices of the graph.

Thus, if for some $\{r_1, \ldots, r_k\} \subseteq \{i_1, \ldots, i_l\}$ we have:
$$(sM\alpha(r_1, j) = 1) \wedge \ldots \wedge (sM\alpha(r_k, j) = 1),$$

then $\{v_{r_1}, \ldots, v_{r_k}\}$ will be an incompatible set of vertices of the graph $CG$ with respect to the column $J_j^\alpha$ of the matrix $sM\alpha$.

For convenience, when describing the algorithm, we will use the following notation:

If the set of vertices $\{v_{r_1}, \ldots, v_{r_k}\}$ of the clean graph is incompatible with respect to the column $J_j^\alpha$, then we will denote it as IncompSet($j$), that is
$$\text{IncompSet}(j) = \{v_{r_1}, \ldots, v_{r_k}\}.$$

Let us first describe the notion of extension of a clean graph, introduced in [8], in order to proceed to the incompatibility elimination procedure.

## 6.2. *Extension of a Clean Graph*

We will describe the extension of the clean graph based on the following conditions:

$\{v_{i_1}, \ldots, v_{i_l}\}$ is the set of vertices of a clean graph $CG(q_1, \ldots, q_l)$.

IncompSet($s_1$), . . ., IncompSet($s_p$) are incompatible sets of vertices in $CG(q_1, \ldots, q_l)$ with respect to the columns $J_{s_1}^\alpha, \ldots, J_{s_p}^\alpha$, respectively, satisfying the following conditions,

- none of the sets IncompSet($s_1$), . . ., IncompSet($s_p$) contains a removable vertex,

- there are ordered pairs that are not included in the description of the replaceability procedure corresponding to the graph $G(q_1, \ldots, q_l)$, let them be
$$(I_{l_1}^\alpha, I_{l_1}^{\bar\alpha}), \ldots, (I_{l_k}^\alpha, I_{l_k}^{\bar\alpha}),$$
such that each column of the set $\{J_{s_1}^{\bar\alpha}, \ldots, J_{s_p}^{\bar\alpha}\}$ contains 1s also in some of the rows $I_{l_1}^{\bar\alpha}, \ldots, I_{l_k}^{\bar\alpha}$.

No other incompatible sets are included in the graph $CG(q_1, \ldots, q_l)$ or all they are eliminated.

We will say that the graph denoted as $ExG(s_1, \ldots, s_p)$ is an extension of the clean graph $CG(q_1, \ldots, q_l)$ associated with the columns $J^\alpha_{s_1}, \ldots, J^\alpha_{s_p}$ if it is constructed by adding new vertices and edges to the graph $CG(q_1, \ldots, q_l)$ as follows:

    - the graph $ExG(s_1, \ldots, s_p)$ includes the graph $CG(q_1, \ldots, q_l)$,

    - the vertices $v_{l_1}, \ldots, v_{l_k}$ corresponding to ordered pairs

$$(I^\alpha_{l_1}, I^{\overline{\alpha}}_{l_1}), \ldots, (I^\alpha_{l_k}, I^{\overline{\alpha}}_{l_k})$$

are considered as additional main vertices for the graph $ExG(s_1, \ldots, s_p)$,

    - new edges and vertices for the graph $ExG(s_1, \ldots, s_p)$ are constructed using the description of the replaceability procedure started from the replacement steps

$$(I^\alpha_{l_1}, I^{\overline{\alpha}}_{l_1}), \ldots, (I^\alpha_{l_k}, I^{\overline{\alpha}}_{l_k})$$

under the decomposition $d_n S$.

The vertices and edges that appeared during this procedure, will not be added to the new graph, if they are already included in it or were removed during any removal procedures.

When forming new edges and vertices, all their parameters are added to the corresponding arrays and matrices.

If no incompatible set satisfies the described conditions, then we assume that the graph $ExG(s_1, \ldots, s_p)$ coincides with the graph obtained as a result of compatibility procedure, and the set $\{J^{\overline{\alpha}}_{s_1}, \ldots, J^{\overline{\alpha}}_{s_p}\}$ is empty.

*The set of columns $\{J^{\overline{\alpha}}_{s_1}, \ldots, J^{\overline{\alpha}}_{s_p}\}$ will also be called the set of graph extension columns.*

*The graph $ExG(s_1, \ldots, s_p)$ will be called final extension for the graph $CG(q_1, \ldots, q_l)$ if there are no more columns that extend the graph.*

In other words, vertices and edges enter the graph $ExG(s_1, \ldots, s_p)$ if and only if they enter the extending graph or appear in the replaceability procedure started from the main replaceability steps $(I^\alpha_{l_1}, I^{\overline{\alpha}}_{l_1}), \ldots, (I^\alpha_{l_k}, I^{\overline{\alpha}}_{l_k})$ under the decomposition $d_n S$.

To describe the extension algorithm, we assume that all the required parameters of the graph are known and the required conditions are satisfied. Note that all these parameters are obtained by executing the described algorithms.

Thus, we assume that:

$\{v_{l_1}, \ldots, v_{l_l}\}$ is the set of vertices of a clean graph $CG(q_1, \ldots, q_l)$,

IncompSet$(s_1)$, ..., IncompSet $(s_p)$ - incompatible sets of vertices in $CG$ with respect to the columns $J^\alpha_{s_1}, \ldots, J^\alpha_{s_p}$, respectively,

$(I^\alpha_{l_1}, I^{\overline{\alpha}}_{l_1}), \ldots, (I^\alpha_{l_k}, I^{\overline{\alpha}}_{l_k})$ - are the ordered pairs not included in the set $R$ of ordered pairs corresponding to the vertices of the clean graph $CG(q_1, \ldots, q_l)$.

$\{J^{\overline{\alpha}}_{s_1}, \ldots, J^{\overline{\alpha}}_{s_p}\}$ – are all columns in the matrix $sM\overline{\alpha}$ such that each of them contains 1s also in some of the rows $I^{\overline{\alpha}}_{l_1}, \ldots, I^{\overline{\alpha}}_{l_k}$.

We proceed as follows:

- we mark vertices $v_{l_1}, \ldots, v_{l_k}$ as already formed vertices,

- we add the numbers of the vertices $v_{l_1}, \ldots, v_{l_k}$ to the array V,

- we mark the vertices $v_{l_1}, \ldots, v_{l_k}$ as additional main vertices for the extending graph. Note that these vertices will be associated with some column $J_s^{\bar{\alpha}} \in \{J_{s_1}^{\bar{\alpha}}, \ldots, J_{s_p}^{\bar{\alpha}}\}$.

- for any $J_{s_i}^{\bar{\alpha}} \in \{J_{s_1}^{\bar{\alpha}}, \ldots, J_{s_p}^{\bar{\alpha}}\}$, we assign the number of main vertices associated with the column $J_{s_j}^{\bar{\alpha}}$, to the element Multiplicity($s_j$) of the array Multiplicity.

- we apply the algorithm $Alg2$ to the vertices $v_{l_1}, \ldots, v_{l_k}$.

### 6.3. *Incompatibility Elimination Procedure in a Graph.*

As it is described in section 6.5 in [8] the incompatibility elimination procedure, also called the compatibility procedure, consists of the sequential application of the removal procedure to the vertices of incompatible sets in the graph. Let's redefine it based on the graph description parameters.

Suppose that $CG(q_1, \ldots, q_l)$ is a clean graph and IncompSet($j$) is an incompatible set of vertices in it with respect to some column $J_j^{\alpha}$ of the matrix $sM\alpha$.

(a) we apply the removal procedure to some vertex $v$ included in IncompSet($j$).

If the vertex $v$ is not removable, we consider the initial graph and apply the removal procedure to another vertex of the set IncompSet($j$).

If a removable vertex is found in the set IncompSet($j$), then the application of the removal procedure to the vertices included in this set is terminated. In this case, we will say that the incompatibility of the set IncompSet($j$) is eliminated and continue to consider other incompatible sets, if any exists.

(b) let IncompSet($j$) be an incompatible set with respect to the column $J_j^{\alpha}$ such that none of the vertices included in it is removable.

In this case we check whether the column $J_j^{\bar{\alpha}}$ contains 1 in the $\bar{\alpha}$-components of some ordered pairs not included in the replaceability procedure.

(b.1) If the column $J_j^{\bar{\alpha}}$ does not contain 1 in the $\bar{\alpha}$-component of any of these ordered pairs, we conclude that the incompatibility of the set IncompSet($j$) is not eliminated.

This means that the graph vertex incompatibility cannot be eliminated, so we terminate the procedure.

(b.2) If the column $J_j^{\bar{\alpha}}$ contains 1 in the $\bar{\alpha}$-components of some ordered pairs not included in the replaceability procedure, then

- we mark $J_j^{\bar{\alpha}}$ as a column which will be considered later,

- we continue to apply the removal procedure to the vertices of the other incompatible set, if any exists.

(c) Assume that, as a result of applying the compatibility procedure to all incompatible sets in the graph, we have the following:

- the incompatibilities of some sets are eliminated,

- the incompatibilities of other sets are not eliminated, let these sets be incompatible with respect to the columns $J_{s_1}^\alpha, \ldots, J_{s_p}^\alpha$ respectively. At the same time each of these columns contains 1s also in the $\bar{\alpha}$-components of some ordered pairs not included in the replaceability procedure.

In this case, we form the graph $ExG(s_1, \ldots, s_p)$ as an extension of the clean graph under consideration, using Description 6.2.

Since new edges and vertices were added to the graph when constructing the graph $ExG(s_1, \ldots, s_p)$, we apply the cleaning procedure to it if useless vertices appear in it.

Also, if the array $M\alpha$ obtained from the decomposition corresponding to the vertices of this graph contains zero elements, then:

- we search for incompatible sets of vertices in the graph,

- we apply the compatibility procedure to these incompatible sets.

We continue the extension of the graph, by the same procedure, exploring the new incompatible sets and adding new main vertices, if they appear during these procedures.

_Definition_ 6.4. We will say that the compatibility procedure in a graph is completed if either the incompatibility of any incompatible set in it or in any of its extension is eliminated, or if an incompatible set is found that cannot be eliminated.

We describe an algorithm, (denoted by $Alg\,4$) that determines whether a given clean graph contains an incompatible set of vertices and, having found it, proceeds to eliminate it by applying the removal procedure to the vertices included in it.

Recall that to obtain an ordered pair of matrices corresponding to the clean graph, only the components of the ordered pairs corresponding to the vertices of this graph are permuted.

Based on the ordered pair of matrices $(sM\alpha, sM\bar{\alpha})$, the algorithm forms the new elements of the arrays $M\alpha$ and $M\bar{\alpha}$.

As defined above, for any $j \in \{1, \ldots, m\}$:

$M\alpha(j)$ shows the number of 1s in the $j$-th column in the matrix obtained from $sM\alpha$ as a result of permutation of the components of the ordered pairs of rows corresponding to the vertices of the clean graph.

$M\bar{\alpha}(j)$ shows the number of 1s in the $j$-th column in the matrix obtained from $sM\bar{\alpha}$ as a result of permutation of the components of the ordered pairs of rows corresponding to the vertices of the clean graph.

In addition, if $M\alpha(j) = 0$ for some $j$, then the graph contains incompatible set of vertices.

To determine these vertices, we look for 1s in the column $J_j^\alpha$ of the matrix $sM\alpha$, since $M\alpha(j) = 0$ due to the displacement of the 1s of the column $J_j^\alpha$. The indices of rows of these 1s will be the indices of incompatible vertices with respect to the column $J_j^\alpha$.

To eliminate the incompatibility of this set, we apply the removal procedure to the vertices included in the set.

We will use the notations for some sets additional arrays and matrices which are forming during the execution the algorithm. They are:

IncompSet(j) is a set of incompatible vertices with respect to the column $J_j^\alpha$ if it exists.

RemovedVertices is an array of indexes of the vertices that are removed during the removal procedure.

GraphExtendVertices is set of vertices that extend the graph.

Col is an array of columns, due to which the vertices appear that extend the graph.

In addition, the algorithm uses some variables:

numIncompSets is the number of incompatible sets in the graph.

numExtendVertices is the number of vertices that extend the graph.

numExtendCols is the number of columns in $sM\bar{\alpha}$ due to which the graph is extended.

zeroColumn is an indicator showing if there exists zero column.

*Alg* 4. *Determination of Incompatible Sets of Vertices and their Elimination*

1. Update arrays $M\alpha$ and $M\bar{\alpha}$
2. **Initialize variables** numIncompSets, numExtendVertices, numExtendCols

3. **While** $j \leq m$:                    // determining an incompatible set of vertices //
4.     **If** $M\alpha(j) = 0$:              // if there is an incompatible set with respect to the column $J_j^\alpha$
5.         numIncompSets := numIncompSets +1

6.     **For** any vertex $v_s$ **until** all vertices are considered **do**:
7.         **If** ($v_s$ is not removed) **and** ($sM\alpha(s, j)$) = 1:
8.             Add $v_s$ to the set IncompSet(j)
9.         **Else**:
10.         **EndIf**;
11.     **EndFor** vertices $v_s$;        //the incompatible set with respect $J_j^\alpha$ is determined //

        // Applying the compatibility procedure to the found incompatible set

12.     **Copy arrays**:  //all parameters that change during the removal procedure are copied//
13.         Multiplicity → Multiplicity1;
14.         Ind → Ind1;
15.         GraphEdges → GraphEdges1;
16.         DisEdges → DisEdges1;

17.　　　**For** any vertex $v_r$ in IncompSet(j) **until** all they are considered **do**:

18.　　　　**If** RP has not yet been applied to the vertex $v_r$:

19.　　　　　　Anc(1) := $v_r$

20.　　　　　　Mark $v_r$ as a vertex to which RP is applied

21.　　　　　　**Call** RP;　　　　　　　　　// apply RP to the vertex IncompSet(j2, j))

22.　　　　　　**If** Rem(Anc(1)) = 1:

23.　　　　　　　　Update the arrays $M\alpha$ and $M\bar{\alpha}$　// since some vertices are removed,
　　　　　　　　　　　　　　the arrays $M\alpha$ and $M\bar{\alpha}$ will be updated //

24.　　　　　　　　j :=1

25.　　　　　　　　**Go to Step** 3;　　　//search for another incompatible set

26.　　　　　　**Else**:　　　　　　　　　　// if $v_s$ is not a removable vertex

27.　　　　　　　**Copy arrays**:　//Restore the arrays since the considered vertex is not
　　　　　　　　　　　　removable//

28.　　　　　　　　Multiplicity1 → Multiplicity;

29.　　　　　　　　Ind1 → Ind;

30.　　　　　　　　GraphEdges1 → GraphEdges;

31.　　　　　　　　DisEdges1 → DisEdges;

32.　　　　　　**EndIf** ;

33.　　　　**Else**:

34.　　　**EndFor** vertex $v_r$;

　　　　// none of the vertices in IncompSet(j) is removable, so we look for 1s that are included in
　　　　the column $J_j^{\bar{\alpha}}$ and are not included in rows participated in the replaceability procedure//

35.　　　additionalOnes = 0

36.　　　**For** ordered pairs $(I_p^\alpha, I_p^{\bar{\alpha}})$ **until** all ordered pairs are considered:

37.　　　　**If** $[(I_p^\alpha, I_p^{\bar{\alpha}})$ is not included in Replaceability Procedure] **and** $[sM\bar{\alpha}(p, j) = 1]$:

38.　　　　　numExtendVertices = numExtendVertices +1

39.　　　　　additionalOnes = additionalOnes +1

40.　　　　　GraphExtendVertices (numExtendVertices) = p

41.　　　　　numExtendCols = numExtendCols +1

42.　　　　　Col(numExtendCols) = j

43.　　　　**Else**:

44.　　　**EndFor** ordered pairs;

45.　　　**if** additionalOnes = 0:

46.　　　　**Output** ("Incompatibility with respect to the column", j, "is not eliminted."
　　　　　　　"$M^\alpha$-covering does not exist.")

47.　　　　**Stop**;

48.　　　**EndIf**;

49.　　**Else**:

50. **EndWhile** j;

51.  **If** numIncompSets = 0:

52.      **Output** ("The graph does not have incompatible sets. $\alpha$-domain of the decomposition
                corresponding to the vertices of the graph is an $M^{\alpha}$-covering for the set ")

53.      **Stop**;

54.  **Else**:

55.      **Go to** *Graph extension procedure*     // *Alg*5

As a result of execution of this algorithm, we will get one of the following outputs:

a) the incompatibility of any incompatible sets is eliminated, so the $\alpha$-domain of the resulting decomposition is a special covering for the set under the given decomposition.

b) there is an incompatible set of vertices in the graph whose incompatibility is not possible to eliminate, so there is no special covering for the set.

c) any incompatible set containing only non-removable vertices is incompatible with respect to the column through which new main vertices appear. So, the algorithm proceeds to extend the graph.

Let's describe an algorithm which extends the graph and searches a special covering in the extended graph.

*Alg*5. *Graph extension procedure*

1.  **If** numExtendVertices = 0:

2.      **Output** ("all incompatible sets are eliminated. So, the $\alpha$-domain of the decomposition
                corresponding to the vertices of the graph is an M alpha covering for the set. ")

3.      **Stop**;

4.  **Else**:

5.      moveCounter = totalVertices

6.      **For** $i = 1$ **to** numExtendVertices **do**:

7.          $l := \text{totalVertices} + i$

7.          V($l$) := GraphExtendVertices ($i$)

8.          Mark $v_{V(l)}$ as already formed vertex

9.          Mark $v_{V(l)}$ as main vertex

10.         **For** j= 1 **to** $m$ **do**:

11.           **If** $sM\bar{\alpha}$(GraphExtendVertices($i$), Col ($j$)) = 1:

12.             multiplicity (Col (j)) = multiplicity (Col (j)) + 1

13.           **Else**:

14.         **EndFor** j;

15.         **EndFor** $i$;

16.     totalVertices = totalVertices + numExtendVertices

17. **Go to** *Alg*2

_Proposition_ 6.5. The number of elementary operations to find all incompatible sets of vertices in the clean graph does not exceed the number $c \times n \times m$.

_Proof._ It is easy to see that based on the ordered pairs of matrices ($sM\alpha$, $sM\bar{\alpha}$), the arrays $M\alpha$ and $M\bar{\alpha}$ can be formed in $c \times n \times m$ elementary operations for some constant $c$.

Let $\{v_{i_1}, \ldots, v_{i_l}\}$ be the set of the vertices of the clean graph.

If for some j $\in \{1, \ldots, m\}$, some vertices are incompatible with respect to the column $J_j^\alpha$ of the matrix $sM\alpha$, then during the execution, the algorithm determines them and forms the set IncompSet(j).

To do this, the algorithm searches for zero element in the array $M\alpha$.

Obviously, all zeros in the array $M\alpha$ can be found in $m$ elementary operations.

Let $M\alpha(j) = 0$ for some j $\in \{1, \ldots, m\}$. Since $M\alpha(j)$ becomes zero due to permutations of components of the ordered pairs corresponding to the vertices of the clean graph, the algorithm looks for 1s in the column $J_j^\alpha$ of the matrix $sM\alpha$.

If for a given value j, the numbers $r_1, \ldots, r_s$ are the indices of the rows of the matrix $sM\alpha$ such that

$$(\{r_1, \ldots, r_s\} \subseteq \{i_1, \ldots, i_l\}) \text{ and } (sM\alpha(r_1, j) = 1, \ldots, sM\alpha(r_s, j) = 1),$$

then for this value of j the algorithm forms the set

$$\text{IncompSet(j)} = \{v_{r_1}, \ldots, v_{r_k}\}.$$

Since $s \leq n$, then this can be done in $c \times n$ elementary operations for some constant $c$.

Obviously, in order to find incompatible sets for all j $\in \{1, \ldots, m\}$ such that $M\alpha(j) = 0$, and form the sets IncompSet(j) for these values of j, we need no more than

$$c \times n \times m$$

elementary operations for some constant $c$.

_Theorem_ 6.6. The number of elementary operations required to complete the compatibility procedure in any pointing graph does not exceed the number

$$c \times m \times n \times (n + m)$$

for some constant $c$.

_Proof_. To complete the compatibility procedure the algorithm will determine the incompatible sets and apply the incompatibility elimination procedure to them.

According to Proposition 6.5 the algorithm finds all incompatible sets of vertices in a clean graph in no more than $c \times n \times m$ elementary operations for some constant $c$.

The compatibility procedure also includes graph extension and, if necessary, the cleaning procedure in the extended graph. That is, we construct new graph and, if useless vertices appear, we eliminate them.

According to Corollaries 4.9.1 and 4.9.2 the number of elementary operations of constructing a graph and applying the cleaning procedure to it, does not exceed the number

$$c \times n(n + m)$$

for some constant $c$.

To have an estimate for the compatibility procedure, we add to this estimate the number of operations that the algorithm does when applying the incompatibility elimination procedure to other sets found.

The following statements are obvious:

a) the number of different incompatible sets in a graph does not exceed the number $m$.

b) the total number of different vertices included in all incompatible sets in any pointing graph does not exceed the number $n$.

c) different incompatible sets in the graph may contain common vertices.

Let the vertex $v$ be included in the incompatible sets

$$\text{IncompSet}(j_1), \ldots, \text{IncompSet} (j_p).$$

c1) if the incompatibility of one of these sets is eliminated by applying the removal procedure to the vertex $v$, then the incompatibility of all these sets will be eliminated simultaneously.

c2) if the incompatibility of one of these sets is not eliminated by applying the removal procedure to the vertex $v$, then the incompatibility none of these sets will be eliminated by applying the removal procedure to the vertex $v$.

c3) the total number of removed edges during the compatibility procedure does not exceed the number $(n - 1) \times m$.

The algorithm execution principle is as follows:

1) if an incompatible set of vertices with respect to some column $J_j^\alpha$ is found, then the algorithm proceeds to eliminate it.

2) the search of the next incompatible set is performed only if the incompatibility of the previous set is eliminated or the column $J_j^{\bar\alpha}$ contains 1s in the $\bar\alpha$-components of ordered pairs not included in the replaceability procedure.

Thus, with this principle, the removal procedure will not be applied to common vertices of different incompatible sets more than once. This means that regardless of the number of incompatible sets containing some common vertex $v$, the removal procedure to this vertex is applied once.

On the other hand, the algorithm performs multiple elementary operations on common vertices when checking whether they have been considered before. The number of these operations for each of them is equal to the number of incompatible sets containing this vertex.

In addition, the algorithm performs $c \times (n + m)$ elementary operations considering any vertex during the compatibility procedure, to find and remove its incoming and outgoing edges, as well as to fix the corresponding vertices that will be removed later.

Taking into account also that the same vertices can occur in more than one incompatible set, we can assert that the number of elementary operations required to complete the compatibility procedure in any pointing graph does not exceed the number

$$c \times m \times n \times (n + m)$$

for some constant $c$. $\nabla$

Let's consider now the procedures P1, P2, P3, P4 of Paragraph 1. For each of these procedures, we have obtained a polynomial estimate of the number of elementary operations.

P1) According to Propositions 4.8, 4.9, 4.9.1, 4.9.2, in order to construct a pointing graph based on some special decomposition of a given set, no more than

$$c \times n \times m$$

elementary operations are required, for some constant $c$.

P2) According to Theorem 5.4, the total number of elementary operations for eliminating all useless vertices in any pointing the graph does not exceed the number

$$c \times n(n + m)$$

for some constant $c$.

P3) According to Theorem 6.6, the number of elementary operations required to complete the compatibility procedure in any pointing graph does not exceed the number

$$c \times m \times n \times (n + m)$$

for some constant $c$.

P4) The graph extending procedure is actually a graph constructing procedure.

Therefore, as in case of P1, the number of elementary operations for constructing an extended pointing graph does not exceed the number

$$c \times n \times (n + m)$$

for some constant $c$.

Combining all these estimates, we can assert that the problem of finding a special covering for a set under the special decomposition of this set is decidable in no more than

$$c \times m \times n \times (n + m)$$

elementary operations, for some constant $c$.

Recall that $N(d_n S)$ is the length of input data of the decomposition $d_n S$, that is, $N(d_n S)$ is the total number of 1s in the matrices $sM\alpha$ and $sM\bar{\alpha}$.

It is easy to see that $m \leq N(d_n S)$ and $n \leq N(d_n S)$, hence it follows that

$$c \times m \times n \times (n + m) \leq c \times (N(d_n S))^3.$$

So, the estimate is polynomial with respect to the number of input data.

_Theorem_ 6.7. Let $d_n S$ be a special decomposition of the set $S$, and $N(d_n S)$ be the length of its input data, then the number of elementary operations required to find out if there is a special covering for the set $S$ under the special decomposition $d_n S$, does not exceed the number

$$c \times (N(d_n S))^3.$$

_Proof._ The proof follows from the estimate of procedures P1, P2, P3, P4. $\nabla$

# 7. Satisfiability of a Boolean Function

The algorithm $SSC$ searches for a special covering for a given set under the given special decomposition of this set based on the pair of matrices $(sM\alpha, sM\bar{\alpha})$ corresponding to this special decomposition.

To study the satisfiability of the Boolean function represented in $CNF$, we consider this function as a set $S(f)$ of its clauses and search for special covering for this set under the special decomposition

$$d_nS(f) = \{(fM_1^\alpha, fM_1^{1-\alpha}), (fM_2^\alpha, fM_2^{1-\alpha}), \ldots, (fM_n^\alpha, fM_n^{1-\alpha})\}.$$

Recall that $d_nS(f)$ is a special decomposition according to Lemma 7.1 in [8].

According to the Theorem 7.2 in [8], if there is a special covering for the set $S(f)$ under some special decomposition $d_nS(f)$, then the function $f$ is satisfiable.

Therefore, we describe an algorithm, denoted by $SatAlg$, that generates a pair of corresponding matrices based on the Boolean formula, represented in conjunctive normal form and apply the algorithm $SSC$ to these matrices.

We are given a Boolean function $f(x_1, \ldots, x_n)$ represented in $CNF$ with $m$ clauses.

We will assume that the clauses of this function are numbered in an arbitrary order, and $1, \ldots, m$ are their numbers. We denote the clauses as $c_1, \ldots, c_m$.

For a function $f$ of $n$ variables and of $m$ clauses represented in $CNF$, we form an $n \times m$ matrix, denoted by $(f)CNF$, as follows:

$(f)CNF(i, j) = $ -1, *if the negative literal $\bar{x}_j$ is included in the clause $c_i$,*

$(f)CNF(i, j) = 0$, *if none of the literals $x_j$ and $\bar{x}_j$ is included in the clause $c_i$,*

$(f)CNF(i, j) = 1$, *if the positive literal $x_j$ is included in the clause $c_i$.*

Obviously, for any $i \in \{1, \ldots, n\}$, the $i$-th row of the matrix $(f)CNF$ is uniquely determined by the clause $c_i$ of the function.

Also, for any $i \in \{1, \ldots, n\}$, the clause $c_i$ of the function is uniquely determined by the $i$-th row of the matrix $(f)CNF$.

So, any Boolean function $f$ represented in $CNF$ is uniquely determined by the matrix $(f)CNF$, and the matrix $(f)CNF$ is uniquely determined by the Boolean function $f$ in $CNF$.

*We will say that the Boolean function $f(x_1, \ldots, x_n)$ in CNF of $m$ clauses is represented by the $n \times m$ matrix $(f)CNF$.*

<u>*Definition*</u> 7.1. The total number of nonzero elements of the matrix $(f)CNF$ will be called the number of input data of the Boolean function represented in $CNF$.

<u>*Proposition*</u> 7.2. The number of input data of any Boolean function represented in $CNF$ is equal to the number of input data of the special decomposition generated by this function.

_Proof_. Recall that the components of the $i$-th ordered pair $(fM_i^\alpha, fM_i^{1-\alpha})$ of the special decomposition generated by the function $f$, for any $i \in \{1, \ldots, n\}$ are formed as follows:

$$fM_i^\alpha = \{c_j \,/\, c_j \in S(f) \text{ and } c_j \text{ contains the literal } x_i^\alpha, \ (j \in \{1, \ldots, m\})\}.$$
$$fM_i^{1-\alpha} = \{c_j \,/\, c_j \in S(f) \text{ and } c_j \text{ contains the literal } x_i^{1-\alpha}, \ (j \in \{1, \ldots, m\})\}.$$

For any $j \in \{1, \ldots, m\}$, we compose the following two sets:

$$L_j^\alpha = \{\, i \,/\, i \in \{1, \ldots, n\} \,\&\, [(f)CNF(i, \ j) = \text{-}1]\},$$
$$L_j^{\overline{\alpha}} = \{\, i \,/\, i \in \{1, \ldots, n\} \,\&\, [(f)CNF(i, \ j) = 1]\}.$$

Obviously, if consider the elements included in these sets as clause numbers of the Boolean function $f$ of $n$ variables represented in $CNF$, then for some $\alpha \in \{0,1\}$, the ordered pair $(L_j^\alpha, \ L_j^{\overline{\alpha}})$ will uniquely determine the ordered pair $(fM_j^\alpha, \ fM_j^{1-\alpha})$ of the special decomposition $d_n S(f)$ generated by the function $f$ and vice versa.

Let's form the matrices $fM\alpha$ and $fM\bar{\alpha}$, for some $\alpha \in \{0,1\}$, based on the matrix $(f)CNF$ as follows:

$$fM\alpha(i, j) = \begin{cases} 1, \ if \ \ j \ \in L_i^\alpha \\ 0, \ if \ \ j \ \notin L_i^\alpha \end{cases} \quad \text{and} \quad fM\bar{\alpha}(i, j) = \begin{cases} 1, \ if \ \ j \ \in L_i^{\overline{\alpha}} \\ 0, \ if \ \ j \ \notin L_i^{\overline{\alpha}} \end{cases}$$

This means that

$$fM\alpha(i, j) = \begin{cases} 1, \ if \ \ (f)CNF(j, \ i) = -1 \\ 0, \ if \ \ (f)CNF(j, \ i) \neq -1 \end{cases} \quad \text{and} \quad fM\bar{\alpha}(i, j) = \begin{cases} 1, \ if \ \ (f)CNF(j, \ i) = 1 \\ 0, \ if \ \ (f)CNF(j, \ i) \neq 1 \end{cases}$$

It is easy to see that the total number of 1s in the ordered pair of matrices $(fM\alpha, fM\bar{\alpha})$ is equal to the total number of 1s in the ordered pair of matrices $(sM\alpha, sM\bar{\alpha})$ for the special decomposition $d_n S(f)$.

On the other hand, it is evident, that the total number of 1s in the ordered pair of matrices $(fM\alpha, fM\bar{\alpha})$ is equal to the nonzero elements of the matrix $(f)CNF$. $\nabla$

Let's describe an algorithm for forming the pair of matrices $(fM\alpha, fM\bar{\alpha})$, which will be the beginning of the algorithm $SatAlg$. However, for convenience, we will use the notation $sM\alpha$ and $sM\bar{\alpha}$ instead of $fM\alpha$ and $fM\bar{\alpha}$, respectively.

Beginning //Input: $(f)CNF$ matrix; Output: the matrices $sM\alpha$ and $sM\bar{\alpha}$ //

1. **For** j = 1 **to** $m$:
2.     **For** $i$ = 1 **to** $n$:
3.         **If** fcnf (j, $i$) = -1:
4.             $sM\alpha$ (i, j) = 1
5.             $sM\bar{\alpha}$ (i, j) = 0
6.         **Elif** fcnf (j, $i$) = 1:
7.             $sM\alpha$ (i, j) = 0
8.             $sM\bar{\alpha}$ (i, j) = 1

9.      **Else**:

10.       $sM\alpha\ (i, j) = 0$

11.       $sM\bar{\alpha}\ (i, j) = 0$

12.      **EndIf**;

13.     **EndFor** $i$;

14.   **EndFor** j;

Obviously, this can be done in $c \times m \times n$ elementary operations for some constant $c$.

After forming of the matrices $sM\alpha$ and $sM\bar{\alpha}$, the algorithm proceeds to $Alg\,1$.

Comparing with the results of Theorem 6.7, we can formulate the main result for the Boolean satisfiability as a Theorem.

_Theorem_ 7.3. Let $f(x_1, \ldots, x_n)$ be a Boolean function represented in $CNF$ with $m$ clauses, and $N(f)$ be the number of input data of this function.

The number of elementary operations required to find out whether the function $f(x_1, \ldots, x_n)$ is satisfiable, does not exceed the number $(N(f))^3$.

_Proof_. The proof follows from Proposition 7.2 and Theorem 6.7. ▽

Recall that as a result of executing the algorithm $SatAlg$, we obtain a special decomposition (the resulting special decomposition) of the set of clauses of the function that satisfies one of the following conditions:

(i) the $\alpha$-domain of this decomposition includes all clauses of the function, which means that all clauses of the function are satisfiable, so the function is satisfiable.

(ii) the $(1- \alpha)$-domain of this decomposition contains clauses that are not $M^{\alpha}$-reachable (Definition 3.1, [8]).

Let's consider the case (ii), which means that as a result of execution of all the procedures associated with the replaceability procedure, non-reachable clauses appeared. That is, we have found clauses that cannot be satisfiable simultaneously with clauses included in the $\alpha$-domain. There is no assignment for variables of the function, that makes all of these clauses satisfiable.

Note that we can move any unreachable clause into the α-domain by permuting the components of the corresponding ordered pair, but this will cause the $\alpha$-domain to lose another clause. At the same time, the assignment, which makes satisfiable all clauses in the resulting $\alpha$-domain, will be different from the previous one.

The resulting estimate for the number of elementary operations refers to the elementary operations required to obtain the resulting special decomposition. Therefore, following the procedure described in 6.14, in [8], it is easy to see that the same estimate will be obtained when searching for the maximum number of satisfiable clauses.

### 7.4. *Space Complexity of the Algorithm SatAlg*

For any Boolean function $f(x_1, \ldots, x_n)$ represented in $CNF$ with $m$ clauses, the algorithm takes as input an $n \times m$ matrix $(f)CNF$.

During the execution the algorithm forms and uses the following arrays and matrices. Matrices:

$sM\alpha(n \times m)$, $sM\bar{\alpha}(n \times m)$, EgdeIn$(m \times n)$ GraphEdges$(n \times n)$, GraphEdges1$(n \times n)$, DisEdge$(n \times m)$, DisEdge1$(n \times m)$, Incomp$(n \times m)$,

One-dimensional arrays:

$M\alpha(m)$, $M\bar{\alpha}(m)$, V$(n)$, Multiplicity$(n)$, Multiplicity1$(n)$, Ind$(n)$, Ind1$(n)$, Anc$(n)$, Gen$(n)$, $Col(m)$, Rem$(n)$, IncompSet$(m)$, GraphExtendVertices$(n)$.

In addition, arrays may be required for marking the main vertices, formed vertices, conjunctive and disjunctive edges of the graph. However, it is easy to see that they will not increase the order of the final estimate of space complexity.

In fact, for any Boolean function, the input data for the algorithm $SatAlg$ consists only of the matrix $(f)CNF$ other parameters are generated during the execution of the algorithm.

Recall that as a length of input data for any Boolean function, we consider the number of non-zero elements of the matrix $(f)CNF$.

The following statements are obvious for any component of the algorithm:

The algorithm operates using only above-mentioned arrays and matrices.

During operation there is no need to extend the amount of space used.

When the pointing graph is extended, new parameters are added to the existing corresponding arrays and matrices.

This means that, the total amount of space required for these arrays and matrices depends only on the numbers $n$ and $m$, that is, on the number of variables and on the number of clauses of the function $f$.

At the same time, it is obvious, that this amount does not exceed the number

$$c \times m \times n$$

for some constant $c$.

Taking into account that $m \leq N(f)$ and $n \leq N(f)$ for any Boolean function of $n$ variables and $m$ clauses represented in $CNF$, we can assert that:

*The total amount of space required for arrays and matrices that uses the algorithm SatAlg searching for the satisfiability of the function f does not exceed the number* $(N(f))^2$.